\documentclass[11pt]{article}

\usepackage{latexsym,amssymb,fullpage}
\usepackage{amsmath,amscd}
\usepackage{latexsym,amssymb,amsmath,amsfonts,amscd,color,amsthm,graphicx}
\usepackage[bold,full]{complexity}
\usepackage{url}

\usepackage[matrix,frame,arrow]{xy}
\usepackage{amsmath}
\newcommand{\bra}[1]{\left\langle{#1}\right\vert}
\newcommand{\ket}[1]{\left\vert{#1}\right\rangle}
\newcommand{\qw}[1][-1]{\ar @{-} [0,#1]}
\newcommand{\qwx}[1][-1]{\ar @{-} [#1,0]}

\newcommand{\gate}[1]{*{\xy *+<.6em>{#1};p\save+LU;+RU **\dir{-}\restore\save+RU;+RD **\dir{-}\restore\save+RD;+LD **\dir{-}\restore\POS+LD;+LU **\dir{-}\endxy} \qw}

\newcommand{\control}{*-=-{\bullet}}

\newcommand{\ctrl}[1]{\control \qwx[#1] \qw}

\newcommand{\Qcircuit}{\xymatrix @*=<0em>}

\providecommand{\abs}[1]{\lvert#1\rvert}

\def\setminus{\smallsetminus}

\def\ie{\textit{i.e.}}
\def\eg{\textit{e.g.}}

\def\al{\alpha}
\def\ba{\beta}
\def\ga{\gamma}
\let\da\delta

\def\rar{\rightarrow}

\def\Rar{\Rightarrow}

\def\msf{\mathsf}

\def\mcl{\mathcal}
\def\mbb{\mathbb}

\newcommand{\mytarg}{*=<0em>{\oplus} \qw}
\newcommand{\inp}[2]{({#1}\;,\;{#2})}
\newcommand{\norm}[1]{\Vert #1 \Vert}

\newtheorem{theorem}{Theorem}[section]\def\TH{\begin{theorem}}\def\HT{\end{theorem}}
\newtheorem{prop}[theorem]{Proposition}\def\PRO{\begin{prop}}\def\ORP{\end{prop}}
\newtheorem{coro}[theorem]{Corollary}\def\COR{\begin{coro}}\def\ROC{\end{coro}}
\def\COR{\begin{coro}}\def\ROC{\end{coro}}
\newtheorem{defi}[theorem]{Definition}\def\DE{\begin{defi}}\def\ED{\end{defi}}
\newtheorem{lemma}[theorem]{Lemma}\def\LE{\begin{lemma}}\def\EL{\end{lemma}}
\newtheorem{algo}{Algorithm}\def\AL{\begin{algo}}\def\LA{\end{algo}}
\newcommand{\AR}[2][c]{$$\begin{array}[#1]{lllllllllllllll}#2\end{array}$$}

\def\MA#1{\left(\begin{matrix}#1\end{matrix}\right)}
\def\EQ#1{\begin{eqnarray}#1\end{eqnarray}}
\def\ket#1{{|}#1\rangle}
\def\bra#1{\langle#1{|}}
\def\ctR{\mathop{\wedge}\hskip-.4ex}

\def\ztwo{{\mbb Z}_2}
\def\ost{\frac1{\sqrt2}}

\def\pit{\frac\pi2}

\def\Cx#1{\cx{#1}{}}
\def\Cz#1{\cz{#1}{}}
\def\cz#1#2{Z_{#1}^{#2}}
\def\cx#1#2{X_{#1}^{#2}}
\def\ei#1{e^{i#1}}

\def\cx#1#2{X_{#1}^{#2}}

\def\mLR#1#2#3#4{{}_{#4}[{M}_{#2}^{#1}]^{#3}}
\def\mR#1#2#3{\mLR{#1}{#2}{#3}{}}
\def\mL#1#2#3{\mLR{#1}{#2}{}{#3}}
\def\m#1#2{{M}_{#2}^{#1}}
\def\M#1#2{{M}_{#2}^{#1}}
\def\et#1#2{E_{#1#2}}

\def\ss#1#2{S_{#1}^{#2}}
\def\mybox#1{\framebox{$#1$}}
\def\CO#1{A_{#1}}

\title{{Parallelizing Quantum Circuits}}

\author{
Anne Broadbent\\\\
\it \small D\'epartement d'informatique \\
\it \small et de recherche op\'erationnelle\\
\it \small Universit\'e de Montr\'eal \\
\small\texttt {broadbea@iro.umontreal.ca}
\and
Elham Kashefi\\\\
\it \small Christ Church College \&\\
\it \small Computing Laboratory\\
\it \small University of Oxford \\
\small \texttt {elham.kashefi@comlab.ox.ac.uk}
}

\begin{document}

\date{April 12, 2007}
\maketitle

\begin{abstract}
We present a novel automated technique for parallelizing quantum
circuits via forward and backward translation to measurement-based
quantum computing patterns and analyze the trade off in terms of depth
and space complexity. As a result we distinguish a class of
polynomial depth circuits that can be parallelized to logarithmic
depth while adding only polynomial many auxiliary qubits. In
particular, we provide for the first time a full characterization of
patterns with flow of arbitrary depth, based on the notion of influencing
paths and a simple rewriting system on the angles of the
measurement. Our method leads to insightful knowledge for
constructing parallel circuits and as applications, we demonstrate
several constant and logarithmic depth circuits. Furthermore, we
prove a logarithmic separation in terms of quantum depth between
the quantum circuit model and the measurement-based model.
\end{abstract}

\section{Introduction and summary of results}

We present a construction for the parallelization of quantum
circuits. Our method gives a formula that computes the exact
decrease in depth that the construction can achieve. This yields
precious insight for the construction of lower-depth quantum
circuits.

The development of parallel  quantum circuits seems almost essential
if we wish to implement quantum algorithms in the near future with
the available technology. Due to decoherence, qubits have a tendency
to spontaneously change their state, hence we can only operate on
them for a very short period of time. Parallel circuits could
maximize the use of these fragile qubits. Note that to obtain
parallelism in the quantum circuit model, we need the ability of interaction with further apart
qubits. Different implementations might put physical
limitations on how far we can apply this ability. However, in some
recent proposals for quantum computing \cite{Pelli97, CZKM97, B00, KLM01,
Nielsen04, BR04, BK04} due  to the construction of the models, the
far apart interaction between qubits have been successfully
demonstrated.

As for theoretical motivation, the study of parallel quantum
algorithms could lead to new results in complexity theory. For
instance, one interesting open question is whether the class of
decision problems solvable in polynomial time,~$\P$, is included in the class
of decision problems solvable in polylogarithmic depth,~$\NC$. Let~$\QNC$ be
the class of decision problems solvable in polylogarithmic depth with a
quantum computer, one can ask similarly whether~$\P$ is included
in~$\QNC$. Finally, Richard Jozsa conjectured that:

{\bf Jozsa Conjecture.}\cite{Jozsa05} \emph{Any polynomial-time
quantum algorithm can be implemented with only $O(\log(n))$ quantum
layers interspersed with polynomial-time classical computations.}

Previous results on parallel quantum circuits include the
parallelization of circuits for the semi-classical quantum Fourier
transform~\cite{GN96}, approximate quantum Fourier
transform~\cite{CW00}, as well as for encoding and decoding quantum
error-correcting codes~\cite{MN02}. These constructions usually
require the use of auxiliary qubits.
The depth complexity of  quantum circuits has also been studied in~\cite{GHP00,GHMP02}.

Our results on parallelizing quantum circuits
(Theorem~\ref{t-cdepthd}) are obtained using the recently proposed
formalism of the measurement-based model for quantum
computation~(MBQC)~\cite{Jozsa05,RB01,Nielsen05,BB06}, an approach
to quantum computing that uses \emph{measurement} as its main
ingredient. A computation in MBQC is usually referred to as a
\emph{pattern} and consists of a round of global operations
(two-qubit gates) to create the required initial multi-qubit
entanglement, followed by a sequence of classically controlled local
operators (single qubit measurements and unitaries). A more formal
definition is given later. We will work in particular within an
algebraic framework for MBQC called the \emph{measurement calculus}
\cite{DKP04}. This novel framework is universal and equivalent in
computational power to the quantum circuit model.\footnote{In this
paper whenever we mention a quantum circuit or a pattern we mean a
uniform family of quantum circuits or patterns, where their
descriptions are given by a classical Turing machine.} Previous
results on the parallelization in the  MBQC include constant-depth
patterns for Clifford unitaries~\cite{RB02} and diagonal
unitaries~\cite{BB06}.

The measurement calculus framework clearly distinguishes between the
quantum and classical depths of a pattern. Informally speaking, the
quantum depth of a pattern is the length of the longest sequence of
dependent commands. The classical depth is the depth of the
classical computation required for the evaluation of the
dependency function of each dependent command. We consider two
transformations that we can apply to patterns without changing their
meaning (the underlying operator that they implement) while never
increasing their depth (and possibly decreasing it): standardization
(Theorem~\ref{t-standardization}) and signal shifting
(Theorem~\ref{t-signal}). Standardization is a rewriting system for
MBQC patterns that pushes all the entanglement operators to the
beginning of the computation, followed by a sequence of the
single-qubit measurements and a final round of local unitaries.
Signal shifting is another rewriting system that translates some of
the quantum depth between measurement operators to classical depth
between the final local unitaries and hence decreases the quantum
depth.

We then develop a method to compute an upper bound on the quantum
depth of a pattern. In order to do so, we use the notion of
\emph{flow} \cite{dk05c}, a graph theoretical tool defined over the
underlying geometry of the initial entanglement state of a pattern.
We further define a construction called an \emph{influencing path},
that allows us to characterize the dependency structure of the
pattern. It is known that a particular set of measurements called
Pauli measurements can be performed independently as the first layer
of measurements \cite{RB01}. Combining this fact about the angles of the
measurement with influencing paths and the signal shifting
procedure, we present an upper bound result on the quantum depth
(Proposition \ref{p-upperdepth}). As for the classical depth, it is
known to be at most logarithmic in the size of the pattern \cite{Jozsa05}.
We give some tighter upper bounds based on the underlying geometry in
Section~\ref{sec:classical depth}.

Our ultimate goal is to decrease the depth of a given circuit,  to
this end we present an automated procedure for the translation of a
circuit  (with $g$ gates) to an MBQC pattern by adding only up
to~$g$ extra auxiliary qubits. Performing standardization and signal
shifting over the obtained pattern might decrease the depth, and we
then translate back the obtained low depth pattern to another
circuit, equivalent to the original circuit but with lower quantum
depth and more auxiliary qubits. This final translation is based on
performing coherent measurements, and therefore the new circuit will
have a  depth equal to the combined quantum and classical depths of
the pattern. Note that since classical computation is cheaper than
quantum computation, one might consider MBQC as a favourable
ultimate architecture for a quantum computer as it keeps the quantum
and classical depth separate. However, this translation forward and
backward to MBQC is interesting from the theoretical point of view
as one can parallelize a circuit automatically and moreover  due to
the simplicity of the translation procedure the pattern depth
characterization of Theorem~\ref{t-depthd} leads to a general
parallelization result for circuits, Theorem \ref{t-cdepthd}.

As already noted, the depth of a pattern is due to the adaptive
measurements and corrections: any given qubit has a fixed set of
measurement outcomes that must be known before a measurement or a
correction command can be performed at that qubit. This set of
measurement dependencies is sometimes called the \emph{backward
cone}~\cite{RB02}. One way of interpreting our main result given by
Theorem~\ref{t-depthd} is that we characterize the backward cone of
any qubit; thus for patterns with flow, we are able to give a method
to easily compute the depth. Moreover our characterization result is
constructive and leads to a novel technique for constructing
parallel patterns and parallel circuits.

In order to demonstrate the power of Theorems~\ref{t-depthd} and~\ref{t-cdepthd}, we present some special cases: depth~2 patterns
(Proposition~\ref{p-depth1}) and depth~2 circuits
(Proposition~\ref{p-cdepth2}). Another application of our results is
Proposition~\ref{p-clifford}, where we show that any polynomial-size
circuit with only Clifford gates can be parallelized to a
logarithmic depth circuit, using a polynomial number of auxiliary
qubits. Using the example of \emph{parity}, we also show a
logarithmic  separation in terms of quantum depth between the
circuit model and the MBQC. Finally, we show how our method can be
used to parallelize a family of polynomial-depth circuits to
equivalent logarithmic depth circuits.

The paper is organized as follows. In
Section~\ref{sec:preliminaries}, we briefly review the MBQC in order
to fix the relevant notation (a more thorough introduction to
quantum computing and MBQC are available in appendices
\ref{Appendix-A} and \ref{Appendix-B}). In Section~\ref{sec:depth},
we define the notion of depth for a pattern in the MBQC, carefully
distinguishing between the preparation, quantum and classical
computation depths. In Section~\ref{sec:standardization and depth},
we show that standardization decreases depth and in
Section~\ref{sec:signalshifting and depth}, we show that signal
shifting also decreases depth. In Section~\ref{s-flowdepth}, we give
upper bounds on the depth of a pattern based on its geometry. In
Section~\ref{s-cir-pattern}, we give a translation from the quantum
circuit model to the measurement-based model and back. Our main
results on characterization of  depth for MBQC patterns and quantum
circuits are given in Section~\ref{s-charact}, where we also present
several applications.

\section{Preliminaries}
\label{sec:preliminaries}

\subsection{Quantum circuit model}

Historically, Richard Feynman was one of the first to suggest that a
computer based on the principles of quantum mechanics could
efficiently \emph{simulate} other quantum systems~\cite{Fey}. David
Deutsch then developed the idea that the quantum computer could
offer a computational advantage compared to the classical computer;
he also defined the \emph{quantum Turing machine}~\cite{Deutsch},
before defining the  \emph{quantum circuit model}~\cite{D89}
 to
represent quantum computations (Deutsch refers to a quantum circuit
as a quantum \emph{network}). It is readily seen that the quantum
circuit model is a generalization of the classical circuit model.

Any unitary operation $U$ can be approximated with a circuit~$C$,
using gates in a fixed universal set of gates (see Appendix
\ref{Appendix-A} or \cite{NC00} for an introduction to quantum
computing). The \emph{size} of a circuit is the number of gates and
its \emph{depth} is the largest number of gates on any input-output
path. Equivalently, the depth is  the number of layers that are
required for the parallel execution of the circuit, where a qubit
can be involved in at most one interaction per layer. In this paper,
we  adopt the model according to which at any given timestep,
a single qubit can be involved in at most one interaction. This
differs from the \emph{concurrency} viewpoint, according to which
all interactions for commuting operations can be done
simultaneously.

\subsection{Measurement-based model}

We give a brief introduction to the MBQC (a more detailed
description is available in Appendix \ref{Appendix-B}
or~\cite{Jozsa05,Nielsen05,BB06,DKP04}). Our notation follows that
of~\cite{DKP04}.

Computations involve the following commands:   1-qubit preparations
$N_i$ (prepares qubit $i$ in state $\ket{+}_i$), 2-qubit
entanglement operators $\et ij:=\ctR Z_{ij}$ (controlled-$Z$
operator), 1-qubit destructive measurements $\M\al i$, and 1-qubit
Pauli corrections $\Cx i$ and $\Cz i$, where $i$, $j$ represent the
qubits on which each of these operations apply, and $\al \in
[0,2\pi)$. Measurement $\M\al i$ is defined by orthogonal
projections onto the state $\ket{+_\al}_i$ (with outcome $s_i=0$) and
the state $\ket{-_\al}_i$ (with outcome $s_i=1$), where
$\ket{\pm_{\al}}$ stands for $\ost(\ket0\pm\ei{\al}\ket1)$.
Measurement outcomes can be summed (modulo~$2$) resulting in
expressions of the form $s=\sum_{i\in I} s_i$ which are called
\emph{signals}.

Dependent corrections are written as $\cx i{s}$ and $\cz i{s}$,
with $\cx i0=\cz i0=I$, $\cx i1=\Cx i$, and $\cz i1=\Cz i$, while
dependent measurements are written as $\mLR{\al} ist$ with \AR{
\mLR{\al} ist = M^\al_i X^s_i Z^t_i = \m{(-1)^s\al+t\pi} i\, . } The
right and left dependency of a measurement are called
\emph{$X$-dependency} and \emph{$Z$-dependency}.

A pattern~$\mcl{P}$ is a finite sequence of commands acting on a
finite set of qubits~$V$, for which~$I\subset V$ and~$O\subset V$
are input and output sets, respectively.  Patterns are executed from
right to left. We assume for the rest of the paper that all the
non-input qubits are prepared and sometimes omit the preparation
commands to be performed at these qubits.

By applying the following \emph{rewrite} rules~\eqref{e-EX}--\eqref{e-MZ} of
the measurement calculus~\cite{DKP04}, we find the \emph{standard}
form of a pattern, which is an ordering of the commands in the
following order: preparation, entanglement, measurement and correction.
\emph{Standardization} is the procedure of applying the rewrite rules until no further rules are applicable.
 \EQ{ \label{e-EX}
\et ij\cx is&\Rar&\cx is\cz js\et ij \\
\label{e-EZ}
\et ij\cz is&\Rar&\cz is\et ij\\
\label{e-MX}
\mLR\al ist\cx i{r}&\Rar&\mLR\al i{s+r}{t}\\
\label{e-MZ} \mLR\al is{t}\cz i{r}&\Rar&\mLR\al i{s}{r+t} }
A pattern which is not in the standard form is called a \emph{wild} pattern.

The \emph{signal shifting
rules}~\eqref{e-signal1}--\eqref{e-signal4} tell us how to propagate
$Z$-dependencies; we refer to \emph{signal shifting} as the
procedure of applying the signal shifting rules until no further
rules are applicable: \EQ{ \label{e-signal1}
\mLR\al ist &\Rar& \ss it  \; \mR\al is \label{ss:eqn1}\\
\label{e-signal2}
\cx js\ss it&\Rar& \ss it \; \cx j{s[(t+s_i)/s_i]}\\
\label{e-signal3}
\cz js\ss it&\Rar& \ss it \; \cz j{s[(t+s_i)/s_i]}\\
\label{e-signal4} \mLR\al jst\ss ir&\Rar& \ss ir \; \mLR\al
j{s[(r+s_i)/s_i]}{t[(r+s_i)/s_i]} } where, $\ss it$ is the signal
shifting command (adding $t$ to $s_i$) and $s[t/s_i]$ denotes the
substitution of $s_i$ with $t$ in $s$.

Dependent commands are essential for universality and the control of
the non-determinism induced by measurements. The following notions
are beneficial for the study of dependency structures of patterns. A
\emph{geometry} $(G,I,O)$ consists of an undirected graph $G$
together with two subsets of nodes $I$ and $O$, called inputs and
outputs. We write $V$ for the set of nodes in $G$, $I^c$, and $O^c$
for the complements of $I$ and $O$ in $V$ and $E_G:=\prod_{(i,j)\in
G}E_{ij}$ for the global entanglement operator associated to $G$
(the graph $G$ is also called the \emph{entanglement graph}
\cite{graphstates}). Trivially, any pattern has a unique underlying
geometry, obtained by forgetting measurements and corrections
commands.

We now give a condition on geometries under which it is possible to
synthesize a set of dependent corrections such that the obtained
pattern is uniformly and strongly deterministic, \ie~all the
branches of the computation are equal, independent of the angles of
the measurements (see Appendix \ref{Appendix-B} for more precise
definitions). Hence we obtain the dependency structure of
measurement commands directly from the geometry, from which we will
get a unified treatment of depth complexity for measurement
patterns. In what follows, $x \sim y$ denotes that  $x$ is adjacent
to $y$ in $G$, $N_{I^c}$ denotes the sequence of preparation
commands $\prod_{i\in I^c} N_i$.

\DE[\cite{dk05c}] A \emph{flow} $(f,\preceq)$ for a geometry $(G,I,O)$ consists of
a map $f:O^c\rar I^c$ and a partial order $\preceq$ over $V$ such
that for all $x\in O^c$:
\begin{itemize}
\item[](i)~~$x \sim f(x)$;
\item[] (ii)~~$x \preceq f(x)$;
\item[] (iii)~~for all $y \sim f(x)$,  $x \preceq y$\,.
\end{itemize} \ED

\begin{figure}
\begin{center}
\includegraphics[scale=0.4]{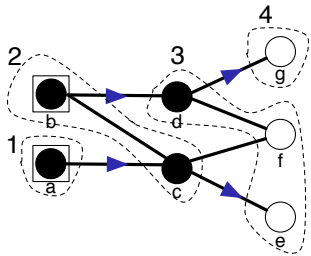}
\caption{An geometry with flow. The boxed vertices are the input
qubits and the white vertices are the output qubits. All the
non-output qubits, black vertices, are measured during the run of
the pattern. The flow function is represented as arcs and the
partial order on the vertices is given by the 4 partition sets. }
\label{flow}
\end{center}
\end{figure}

Figure~\ref{flow} shows  a geometry together with a flow, where $f$
is represented by arcs from~$O^c$ (measured qubits, black vertices)
to~$I^c$ (prepared qubits, non boxed vertices). The associated
partial order is given by the labelled sets of vertices. The
coarsest order $\preceq$ for which~$(f, \preceq)$ is a flow is
called the \emph{dependency order} induced by~$f$ and its depth (4
in Figure~\ref{flow}) is called \emph{flow depth}.

\TH[\cite{dk05c}] \label{t-flow} Suppose the geometry $(G,I,O)$ has
flow~$f$, then the pattern: \AR{ \mcl
P_{f,G,\vec\al}&:=&\prod\limits_{i\in O^c}{\!\!\!}^{\preceq}\,\,\,
\Big(\cx{f(i)}{s_i}\prod\limits_{\substack{k \sim f(i) \\ k \ne
i}}\cz{k}{s_i}\m{\al_i}i\Big) E_G } where the product follows the
dependency order $\preceq $ of $f$,  is uniformly and strongly
deterministic, and realizes the unitary embedding: \AR{
U_{G,I,O,\vec\al}&:=& {2}^{|O^c|/2}\;\Big(\prod\limits_{i\in
O^c}\bra{{+_{\al_i}}}_i \;\Big)\, E_G } \HT

The Flow Theorem (Theorem~\ref{t-flow}) plays an important role in
our discussion of depth complexity in the following sections.
If the underlying geometry of a pattern has flow and pattern commands
sequence is constructed as given by the Flow Theorem, we call
this pattern a \emph{pattern with flow}. Note that the Flow Theorem
tells us how to perform dependent corrections according to the
flow function $f$: when qubit $i$ is measured, its neighbour according to the
flow, $f(i)$, receives the $X_{f(i)}^{s_i}$ correction, while all
the neighbours $k$ of $f(i)$ (independently of the flow and except
$i$), receive a~$Z_k^{s_i}$ correction. We can apply the rewrite
rules of equations~\eqref{e-MX} and~\eqref{e-MZ} to propagate these
dependent corrections to the end and obtain a standard form for the
pattern with flow: \EQ{ \label{e-dependpattern} \prod\limits_{i\in
O}{\!\!\!} \,\,\, \cx i {s_{f^{-1}(i)}} \cz i {\sum_{j: f(j)\sim i}
s_j} \prod\limits_{i\in O^c}{\!\!\!}^{\preceq}\,\,\, \mLR {\al_i} i
{s_{f^{-1}(i)}} {\sum_{j: f(j)\sim i} s_j}E_G } where $f^{-1}(i)$ is
well defined since by construction $f$ is an one-to-one function. If
$f^{-1}(i)$ is empty we ignore the
term $s_{f^{-1}(i)}$, that means the measurement at qubit $i$ has no
$X$-dependency.

Given a geometry on $n$ vertices with $\abs{I} = \abs{O}$, one can
efficiently \ie~in $O(\poly(n))$) time, find its unique flow if it
exists ~\cite{Beaudrap06,BDK06} and the obtained pattern implements
a unitary operator.

\section{Depth complexity for measurement patterns}

\label{sec:depth}

In this section, we give a definition for the preparation depth and
give its exact value. We also give a definition for the quantum
computation depth of a pattern. Another type of depth exists for a
pattern, this is  the \emph{classical} depth and will be addressed
in Section~\ref{sec:classical depth}.

First, we focus on the notion of depth complexity for a standard
pattern, which we then extend to wild patterns. There are two parts
of a standard pattern computation that contribute to its depth: the
\emph{preparation} phase, which is the work required to prepare the
entangled state (the $N$ and $E$ commands), and the \emph{quantum
computation} phase, which is the work required to perform the
measurements and corrections (the adaptive $M$ and $C$ parts). The
total depth of a pattern in standard form is the sum of the depths
of the preparation and computation parts, which we address now
separately.

\subsection{Preparation depth}
\label{sec:preparation depth complexity}

As already mentioned, for any pattern $\mcl P$ with computational
space $(V,I,O)$ one can associate an underlying geometry $(G,I,O)$
defined by forgetting the measurement and correction commands.  The
entangled state corresponding to this geometry is defined by
preparing the input qubits in the given arbitrary states and all
other qubits in the $\ket +$ state and applying a $\ctR Z$ on all
qubits $i$ and~$j$ that are adjacent in the entanglement graph~$G$.
We give below an exact value for this depth, in terms of
$\Delta(G)$, the maximum degree of~$G$. A similar result also
appeared in~\cite{MS04}.

\LE \label{lemma:preparation} The preparation depth for a given
entanglement graph~$G$, is either~$\Delta(G)$ or~$\Delta(G) +1$. \EL

\begin{proof}
At each timestep, a given qubit can interact with at most one other
qubit. In terms of the entanglement graph, this means that at each
timestep, a given node can interact with at most one of its
neighbours. Assign a colour to each timestep and colour the edge in
the entanglement graph~$G$ accordingly. With this view, the entire
preparation corresponds to an edge colouring of the entanglement
graph. By Vizing's theorem~\cite{Diestel}, the edge-chromatic number
of~$G$, $\chi'(G)$ satisfies $\Delta(G) \leq \chi'(G) \leq \Delta(G)
+1$. \end{proof}

It is known  that a special type of entanglement graph, the
two-dimensional grid (called \emph{cluster state}), is universal for
the measurement-based model. The cluster state is a \emph{bipartite
graph} and hence by K\"onig's theorem~\cite{Diestel}, its
edge-chromatic number is~$\Delta(G)$, hence from the above Lemma we
conclude that any unitary can be implemented with a cluster state
that can be prepared in depth~4.
 This however, might
force the use of extra auxiliary qubits. The precise tradeoff will
be given in Section~\ref{s-cir-pattern} where we deal with the
translation between the circuit model and measurement-based model to
reduce depth.

\subsection{Quantum computation depth}
\label{ss-compdepth}

The \emph{quantum computation} depth of a pattern, or just
\emph{quantum} depth for short is the depth in the execution of the
pattern that is due to the dependencies of measurement and
correction commands on previous measurement results (this is also  called the \emph{causality depth}).  Given
a pattern in standard form, it is easy to calculate its quantum
computation depth from its \emph{execution} digraph given below.

\DE  \label{d-exec} The \emph{execution digraph}~$R$ for a
pattern~$\mcl{P}$ in standard form has $V$ as node-set. Let the
\emph{domain} of a signal  be the set of qubits on which it
depends. The arcs of $R$ are constructed in the following way:
\begin{enumerate}
\item \label{step:1}Draw an arc from $i$ to $j$ whenever ${_t}[M_j]^s$ appears in the pattern, with~$i$ in the domain of~$s$ or~$t$.
\item \label{step:2} Draw an arc from $i$ to $j$  whenever $X_j^s$ or $Z_j^s$ appears in the pattern, with~$i$ in the domain of~$s$.
\end{enumerate}
\ED

We refer to the nodes of \emph{in-degree} zero in $R$ as
\emph{start} nodes. Similarly, the nodes of \emph{out-degree} zero
in $R$ are called \emph{end} nodes. If there is an arc from~$i$
to~$j$ in the execution digraph, we say that~$j$ \emph{depends} (or
has a \emph{dependency}) on~$i$. As a consequence of the
definiteness condition (see Appendix \ref{Appendix-B}), the graph of
any pattern is acyclic and hence we can give the following
definition for the quantum computation depth:

\DE \label{d-depth} Let $\mcl{P}$ be a pattern in standard form. The
\emph{quantum computation depth} for $\mcl{P}$ is the number of
vertices on the longest directed path between a start and end node
in the execution digraph. We call such a longest path a
\emph{critical path}. \ED

\begin{figure}
\begin{center}
\includegraphics[scale=0.4]{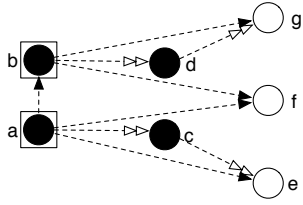}
\caption{The execution digraph for the standard pattern of Equation
\eqref{e-standardpattern}. A white double arrowed arc represents an
$X$-dependency, and a black arrowed arc a $Z$-dependency. The
$X$-dependency arcs correspond to edges of the underlying geometry,
however this is not the case for the $Z$-dependency arcs.}
\label{execution}
\end{center}
\end{figure}

As an example, consider again the geometry given in
Figure~\ref{flow}, one can write a uniformly and strongly
deterministic pattern on this geometry using the Flow Theorem that
can be rewritten in the following standard form:

\EQ{ \label{e-standardpattern} \cz g {s_b}\cx g {s_d}\cz f {s_b}\cz
f {s_a}\cz e {s_a}\cx e{s_c}\mR \da d{s_b} \mR \ga c{s_a} \mL \ba b
{s_a} \m \al a E_G\, , } where $G$ is the entanglement graph
corresponding to the geometry of Figure~\ref{flow}.  Following
Definition~\ref{d-exec}, the execution digraph for the above pattern
is given in Figure~\ref{execution}. As said before (see also
Appendix \ref{Appendix-B}, equations \eqref{e-Xact} and
\eqref{e-Zact}), there are two types of dependent measurements
defined by $X$ and $Z$-dependencies, that are represented with
different arrows in Figure \ref{execution}. The longest path in the
execution graph is $abdg$, hence from Definition \ref{d-depth}, the
pattern depth is~4.

It is trivial that for standard patterns with flow, the quantum
computation depth is the same as the flow depth. However for wild
patterns, the quantum computation depth cannot be dissociated from
the preparation depth (the~$E$ commands being interspersed within
the pattern). In order to define this combined preparation and
quantum depth, we define the  execution digraph  in a similar way as
Definition~\ref{d-exec}, but we add the~$E$ commands to the
execution digraph. Then the \emph{depth} of a wild pattern is the
longest path in the execution digraph \emph{except} that we allow a
sequence of~$E$ commands to be parallelized, the depth of such a
sequence being given by the results of Section~\ref{sec:preparation
depth complexity}.

\section{Standardization reduces depth}
\label{sec:standardization and depth}

In this Section, we refer to the combined preparation and quantum
depth of a pattern as its \emph{depth}.  Intuitively, we would
expect that standardization could only potentially decrease this
depth.
 This is
because by standardizing, we benefit from the fact that there is a
single entanglement graph to consider. Also, corrections are
propagated to the end and applied only on output qubits, hence
potentially fewer operations are needed. On the other hand,
standardization creates dependent measurements.  The following
theorem (which is general and independent of the flow construction)
confirms all these observations. Let $ \mcl{P}\Rightarrow ^\star
\mcl{P}'$ denote the fact that $\mcl{P'}$ is obtained from $
\mcl{P}$ by applying a finite sequence of rewrite rules given by
equations~\eqref{e-EX}--\eqref{e-MZ}.

\TH\label{t-standardization} Whenever  $\mcl{P}\Rightarrow ^\star
\mcl{P'}$ where $ P'$ is in standard form,  the depth of~$\mcl{P'}$
is less than or equal to the depth of~$\mcl{P}$. \HT

\begin{proof}
First, we apply the free commutation rules on those commands
operating on disjoint sets of qubits (see Appendix \ref{Appendix-B})
to~$\mcl{P}$ to obtain~$\mcl{P^*}$ as a sequence of standard
patterns. The depth as defined by the execution digraph is not
affected by this procedure, and $\mcl{P^*}$ depth is actually the
sum of the depths of its standard parts. Thus it is sufficient to
show that standardization of a wild pattern $\mcl{P}$, containing
two parts in standard form, say $\mcl{P}=C^2M^2E^2C^1M^1E^1$ (where
some or all of the parts may be empty), does not increase its depth.
The theorem then follows by induction.

\emph{Step 1. (The $E$'s)}

We show how the re-writing rules are used to bring the pattern
$\mcl{P}=C^2M^2E^2C^1M^1E^1$ to $\mcl{P'}=C^2M^2C^{1'}M^1E^2E^1$ and
that by doing so, the depth of $\mathcal{P'}$ is no greater than
that of $\mcl{P}$. The result holds trivially, if $E^2$ is empty.
Otherwise, for every command $E_{ij} \in E^2$, commute it to the
right-hand side of the pattern by doing the following:

\begin{enumerate}

\item If $C^1$ contains $Z_i$ or $Z_j$ corrections, but no $X_i$ or~$X_j$ corrections,
 we apply the rewriting rule $E_{ij}Z_i^s \Rightarrow Z_i^s
E_{ij}$ and hence the depth does not increase. We then  complete the
commutation by applying the free commutation rules.

\item If $C^1$ contains $X_i$ or $X_j$, then the rewriting rule
$E_{ij}X_i^s \Rightarrow X_i^sZ_{j}^sE_{ij}$ applies. Here, the
command $X_i^s$ has an $s$ dependency, which obviously cannot
contain~$i$ or~$j$, since these qubits haven't been measured yet.
Since $X_i^s$ and $Z_i^s$ do not depend on each other, the addition
of the extra correction does not contribute to the depth.We then
complete the commutation by applying the free commutation rules.
\end{enumerate}

Finally, consider the entanglement graph for $E^1E^2$. Since
$\chi'(E^1 \cup E^2) \leq \chi'(E^1) + \chi'(E^2)$, clearly, the
preparation  depth for $E^1 \cup E^1$ cannot be any greater than the
depth of preparation for $E^1$ plus~$E^2$. Also, as an extra bonus,
since $E_{ij}$ is self-inverse, if $E^1$ and $E^2$ happen to have
common commands, they will cancel out.

\emph{Step 2. (The $M$'s)}

We will show how the free commutation rules and the re-writing rules
are used to bring the pattern $\mcl{P'}=C^2M^2C^{1'}M^1E^2E^{1}$ to
its standard form $\mcl{P''}= C^2C^{1''}M^{2'}M^1E^2E^{1}$  and that
doing so, the depth of $\mcl{P''}$ is no greater than that of
$\mcl{P'}$.

\begin{enumerate}
\item \label{step1}
Consider a command~$_t[M_i^\alpha]^s \in M^2$. If $C^{1'}$ does not
contain any commands acting on qubit~$i$, then~$_t[M_i^\alpha]^s$
freely commutes in~$C^{1'}$, hence we have
commuted~$_t[M_i^\alpha]^s$ to the right-hand side of~$C^{1'}$, and
clearly the sum of the depths of the patterns $C^2M^2$ and
$C^{1'}M^1E^2E^{1}$ is greater than or equal to the depth of the
pattern $C^2C^{1'}M^{2}M^1E^2E^{1}$.

\item \label{step2} Otherwise, we apply the rewrite rules of
equations~\eqref{e-MX} and \eqref{e-MZ}.
 In
the execution digraph,  both of these rules correspond to an
\emph{edge contraction}~\footnote{The \emph{contraction} of edge
$e=(x,y)$ of a graph $G$ is obtained by contracting the edge $e$
into a new vertex, $v_e$, which becomes adjacent to all the former
neighbours  of~$x$ and~$y$. This definition also applies to
digraphs.}. The depth of the execution digraph obviously does not
increase by contracting any of its edges. Hence, after the
applicable edge contractions, we can commute the commands as above,
yielding a pattern of smaller or equal depth. \qedhere
\end{enumerate}
\end{proof}

Theorem \ref{t-standardization} shows us that  in order to improve
the parallel run-time of the pattern, we should implement the
standard  form of the pattern. We also know that standardization can be performed in polynomial time~\cite{DKP04}.  Thus in the
remainder of the paper, we will only consider standard patterns,
which also allows us to consider the preparation depth separately
from the quantum computation depth. Combined with
Theorem~\ref{t-signal} of the next section, we note that the most
efficient form for the implementation of a pattern is also the
signal-shifted form.

\section{Signal shifting reduces depth} \label{sec:signalshifting and depth}

The signal shifting rules (Equations
\eqref{e-signal1}--\eqref{e-signal4}) tell us how we can push
the~$Z$ dependencies of a pattern all the way to the end, which can
then decrease the quantum  depth of a pattern in standard form. We
first present an example and then prove the result, which is also
general and independent of the flow construction.

\begin{figure}
\begin{center}
\includegraphics[scale=0.4]{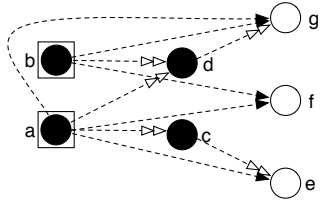}
\caption{The execution digraph for the standard pattern of Equation
\eqref{e-standardpattern} after signal shifting. All the
$Z$-dependencies are pushed to the end and the depth of pattern is now
only~3.} \label{newexecution}
\end{center}
\end{figure}

{\bf Example.} Consider the standard pattern given in
Equation~\eqref{e-standardpattern}. After, signal shifting, we
obtain the following equivalent pattern: \AR{
& & \cz g {s_b}\cx g {s_d}\cz f {s_b}\cz f {s_a}\cz e {s_a}\cx e{s_c}\mR \da d{s_b} \mR \ga c{s_a} \mybox{\mL \ba b {s_a}} \m \al a E_G\\
&\Rar_{\mbox{Eq.} (\ref{e-signal1})}&  \cz g {s_b}\cx g {s_d}\cz f {s_b}\cz f {s_a}\cz e {s_a}\cx e{s_c}\mR \da d{s_b} \mR \ga c{s_a} \ss b {s_a}\m \ba b \m \al a E_G\\
&\Rar&  \cz g {s_b}\cx g {s_d}\cz f {s_b}\cz f {s_a}\cz e {s_a}\cx e{s_c}\mybox{\mR \da d{s_b} \ss b {s_a}}\mR \ga c{s_a}\m \ba b \m \al a E_G\\
&\Rar_{\mbox{Eq.} (\ref{e-signal4})}&  \cz g {s_b}\cx g {s_d}\mybox{\cz f {s_b}\ss b {s_a}}\cz f {s_a}\cz e {s_a}\cx e{s_c} \mR \da d{s_b+s_a} \mR \ga c{s_a}\m \ba b \m \al a E_G\\
&\Rar_{\mbox{Eq.} (\ref{e-signal3})}& \mybox{ \cz g {s_b}\ss b {s_a}}\cx g {s_d}\cz f {s_b+s_a}\cz f {s_a}\cz e {s_a}\cx e{s_c} \mR \da d{s_b+s_a} \mR \ga c{s_a}\m \ba b \m \al a E_G\\
&\Rar_{\mbox{Eq.} (\ref{e-signal3})}& \cz g {s_b+s_a}\cx g {s_d}\cz f {s_b}\cz e {s_a}\cx e{s_c} \mR \da d{s_b+s_a} \mR \ga c{s_a}\m \ba b \m \al a E_G\\
} where boxes represent terms  to be rewritten. Now in the new
execution digraph  (Figure \ref{newexecution}) the longest path has
only three vertices, hence signal shifting has decreased the depth
by~one.

The following theorem states that, in general, signal  shifting does
not increase the depth of a standard pattern. As we have seen above,
it can sometimes decrease~it.

 \TH \label{t-signal} Signal shifting
for a standard pattern does not increase the depth. \HT

\begin{proof}
Let $\mcl{P}$ be a pattern in standard form and suppose that
$\mcl{P}$ includes a command~$_t[M_i^\alpha]^s$ which generates
the signal shifting command~$S_i^t$. Let $\mcl{P'}$ be the pattern that corresponds to the
pattern after the signal $S_i^t$ has been shifted. Let $D$ be the
execution digraph for $\mcl{P}$ and let $D'$ be the execution
digraph for~$\mcl{P'}$. We want to show that the length of a
critical path of~$D'$ is no greater than the length of a critical
path of~$D$.

Suppose that the domain of $s$ in $_t[M_i^\alpha]^s$ is $s_1, s_2,
\ldots ,s_n$ and  that the domain of $t$ is $t_1, t_2, \ldots, t_m$.
Consider all the commands that appear \emph{after}
$_t[M_i^\alpha]^s$ in $\mcl{P}$ and that have an $i$ dependency;
denote these commands $C_{a_1}^i, C_{a_2}^i, \ldots, C_{a_k}^i$
(these are either corrections or measurements). We will show that
the depth does not increase when we shift the signal~$S_i^t$ passed
all the~$C_a^i$'s.

Consider the arcs in $D$ that represent the dependencies between the
measurement ${}{_t}[M_i^\alpha]^s$ and measurements of qubits $t_1,
t_2, \ldots, t_m$; these are the arcs $t_j i$ (for $j=1 \ldots m$)
and we will call these the  \emph{old} arcs. So the old arcs
represent $Z$-dependencies for the ${}{_t}[M_i^\alpha]^s$
measurement. These are precisely those that create signal shifting
commands, since ${}{_t}[M_i^\alpha]^s \Rightarrow
S_i^t\,[M_i^\alpha]^s$.

Now consider the arcs in $D'$ that represent the dependencies
between the measurement of qubit $t_j$, $M_{t_j}$ $(j=1 \ldots m)$
and the measurements and corrections that have an $i$ dependency,
$C_{a_x}^i$ ($x=1 \ldots k$). We call these arcs \emph{new} arcs
since they represent the new dependencies created by $S_i^{t_1},
S_i^{t_2}, \ldots, S_i^{t_n}$ by the signal shifting rules given in
equations~\eqref{ss:eqn1}--\eqref{e-signal4}.

Indeed, when we apply signal shifting to~$\mcl{P}$, we get rid of
all the dependencies represented by old arcs, yet we add all the
dependencies represented by new arcs. These are the only differences
between the execution digraphs $D$ and $D'$.

If all new arcs are already in $D$ (this could be the case if all
the dependencies were present before signal shifting), the graph
$D'$ cannot have a longer critical path than $D$ and we are done.
Otherwise, suppose for a contradiction that the length of a critical
path in $D'$ is greater than the length of a critical path in $D$.
Since $D'$ differs from $D$ only by the removal of all the
\emph{old} arcs and the addition of all the \emph{new} arcs, the
only way for $D'$ to have a longer critical path than $D$ would be
for this critical path to include a \emph{new} arc, say $t_j a_k$
 (obviously, the removal of the \emph{old} arcs in $D'$
cannot contribute to a longer critical path.) But if such a critical
path exists in $D'$, then $D$ admits a longer critical path, namely
the same critical path in $D'$, but with arcs $t_j i$  and $i a_k$
instead of arc $t_j a_k$. This contradiction proves our claim.
\end{proof}

There is a tradeoff when we perform signal shifting, as it can
increase the classical depth. However, as we will show later, the
classical depth is at most $O(\log(n))$  at each layer, where~$n$ is
the number of measured qubits  (Proposition~\ref{t-cdepth})
and hence the tradeoff is beneficial, especially from the
point of view that classical computation is cheap and reliable,
compared to quantum computation that is expensive, error-prone and
subject to  decoherence.

\section{Flow and pattern depth} \label{s-flowdepth}

While the results of Sections~\ref{sec:standardization and depth}
and~\ref{sec:signalshifting and depth} deal with the depth of
\emph{any} pattern, we now focus our attention on the depth of
patterns with \emph{flow}. The flow condition is
sufficient but not necessary for determinism~\cite{g-flow}, however
the class of  patterns with flow is still an interesting class of
patterns, as it is universal for quantum computing, closed under
composition and more importantly our translation from circuits to
patterns in Section~\ref{s-cir-pattern} always yields a pattern
with flow. For the rest of the paper we consider only patterns with
flow.

It is known that for patterns with flow and equal input and output
number of qubits, \ie~those implementing a unitary operator, the
flow, if it exists, is unique~\cite{Beaudrap06}.  From this, we
obtain an upper bound on the quantum computation depth directly from
the underlying geometry. We first define an important notion of
\emph{influencing paths} for geometries.

\begin{defi} Let $(f,\preceq)$ be the flow of a geometry
$(G,I,O)$. Any input-output path in $G$ that  starts with a flow
edge and has no two consecutive non-flow edges  is called an
\emph{influencing path}. \end{defi}

The following are examples of several influencing paths in the
geometry  with flow of Figure \ref{flow} in
Section~\ref{ss-compdepth}: \AR{ ace, acf, acbdf, acbdg\,. }

\begin{prop}
\label{prop:depend-ipaths} Let $a$ and $b$ be two qubits in a
standard pattern with flow.
If~$b$ depends on~$a$, then~$a$ appears before~$b$ on a
common influencing path, and this hold both before and after signal shifting.
\end{prop}

\begin{proof}
This is a consequence of the Flow Theorem. Recall that before signal
shifting, a measurement at a qubit~$j$ is $X$-dependent on the
result of a measurement at another qubit~$i$ if and only if $j=f(i)$
that is, a flow edge between qubits~$i$ and~$j$. Also a measurement
at a qubit~$k$ is $Z$-dependent on the result of a measurement at
another qubit~$i$ if and only if $j=f(i)$ and~$k$ is connected
to~$j$, that is a non-flow edge between qubits~$j$ and~$k$ connected
to a flow edge between qubits~$i$ and~$j$. Therefore signal shifting
creates new dependencies only through influencing paths. Hence if
qubit~$b$ depends on qubit~$a$, it is either via a direct $X$ or $Z$
dependency or due to a sequence of dependencies after signal
shifting, in all the cases $a$ and $b$ must be on a common
influencing path.
\end{proof}

Proposition~\ref{prop:depend-ipaths} tells us that in order to
compute the quantum depth of a standard pattern with flow (to which
we either have or haven't applied signal shifting), it suffices to
consider the depth along influencing paths. Note that after signal
shifting, $Z$-dependencies coming from the non-flow edges on an
influencing path  no longer contribute to the pattern depth, as the
dependencies that they represent are pushed to the final correction
on an output qubit. On the other hand, signal shifting can create
new $X$-dependencies. The following proposition presents an upper bound on the effect
of signal shifting on the pattern depth.

\PRO\label{c-depth1} Let $\mcl P$ be a pattern with flow where
standardization and signal shifting have been performed. Then the
maximum number of flow edges, minus the number of the non-flow edges
on such path (maximum taken over all possible influencing paths),
plus~1 is an upper bound for the depth of the pattern. \ORP
\begin{proof}
We show that for any influencing path, its number of flow edges
minus the non-flow edges gives an upper bound on its depth. Then, by
Proposition~\ref{prop:depend-ipaths}, it suffices to find the
largest number of flow edges along any influencing path in order to
have an upper bound on the depth. We add~$1$ to this depth since the
depth is the number of vertices of such path, and not the number of
edges.

\begin{figure}
\begin{center}
\includegraphics[scale=0.4]{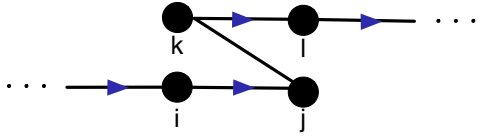}
\caption{Part of an influencing path where two sequence of consecutive flow edges are connected with a non-flow edge.}
\label{influpath}
\end{center}
\end{figure}

Consider an influencing path~$I$. The flow edges represent
$X$-dependencies hence each flow edge in a sequence of consecutive
flow edges contributes to the depth along~$I$. Now, consider a
configuration with a non-flow edge as shown in
Figure~\ref{influpath}. Before signal shifting, the dependent
measurements on  qubits $i$,$j$, $k$ and $\ell$ are given as follows
(see Equation \eqref{e-dependpattern}) where $A,B$ and $C$ stand for
general signals not including $s_i , s_j , s_k$ and $s_\ell$ \AR{
\dots \mLR {\al_\ell} \ell {s_k}{D} \, \mLR {\al_k} k {B}{C+s_i} \,
\mLR {\al_j} j {s_i}{A} \dots } and after signal shifting we have
\AR{ &&\cdots \mLR {\al_\ell} \ell {s_k}{D} \mLR {\al_k} k
{B}{C+\mybox{s_i}} \mLR {\al_j} j {s_i}{A}
\cdots\\
&\Rar& \cdots \mybox{\mLR {\al_\ell} \ell {s_k}{D} \ss k {s_i}}\mLR
{\al_k} k {B}{C} \mLR {\al_j} j {s_i}{A}
\cdots\\
&\Rar& \cdots \ss k {s_i}\mLR {\al_\ell} \ell {s_k+s_i}{D} \mLR
{\al_k} k {B}{C} \mLR {\al_j} j {s_i}{A}
\cdots\\
} Therefore qubits $j$ and $\ell$ are in the same layer. In other
words, after signal shifting,  the first flow edge after every
non-flow edge does not contribute to the depth of the pattern. Also,
any new $X$-dependency created with signal shifting will not
increase the depth.
 Hence from the
total number of the flow edges on an influencing path we need to
subtract the number of non-flow edges.
\end{proof}

So far we have not taken into account the information about the
angles, which is why our bounds are not tight. We first describe the
effect of the Pauli measurements on depth. The following identities are useful
\begin{align}
\label{e-yrule}
\m {\frac{\pi}{2}}i \cx is &= \m{\frac{\pi}{2}}i \cz is
\\ \label{e-xrule}
\m 0i \cx is &= \m 0i
\end{align}

According to Equation~\eqref{e-yrule}, when a qubit~$i$ is measured
with angle~$\pit$ (Pauli~$Y$ measurement), then any $X$-dependency
on this qubit is the same as a $Z$-dependency. But after signal
shifting, this  $Z$-dependency does not directly contribute to the
depth and hence we might obtain a smaller depth. Furthermore, there
exists a special case where if qubit~$i$ is not an input qubit and
also not the flow image of any other vertex ($ \forall j:
i\not=f(j)$) and qubit~$i$ is measured with~$\pit$, then one can
permit in the Flow Theorem, to have $f(i)=i$
and hence we will have one less flow edge~\cite{dk05c}. This allows
an influencing path to have a loop edge on this particular vertex
measured with Pauli~$Y$ and hence the influencing path will not
start with an input qubit. In the rest of the paper, we consider
only this extended notion of influencing path that takes into
account the angles of measurement. When we want to emphasize this
extended definition, we will refer to \emph{Pauli influencing
paths}.

According to Equation~\eqref{e-xrule}, another special case is when
qubit~$i$ is measured with angle~$0$ (Pauli~$X$ measurement), then
any $X$-correction on qubit~$i$ can be ignored and in fact qubit~$i$
can be put at the first level of measurement. Consequently, again
the flow depth can become smaller. By adding
equations~\eqref{e-yrule} and~\eqref{e-xrule} to the Flow Theorem,
the proof still works~\cite{dk05c} and we get a potential
improvement on the depth complexity. We refer to this procedure as
\emph{Pauli simplification}. Another way of realizing these special
cases is that after signal shifting, the Pauli measurements become
independent measurements and hence can all be performed at the first
level of the partial order.  Hence in computing the depth of a
pattern with flow after signal shifting is performed, one should
disregard the Pauli measurements:

\PRO \label{p-upperdepth} Let $\mcl P$ be a pattern with flow where
standardization, Pauli simplification and signal shifting  have been
performed. Let $I_i$ be a Pauli influencing path of~$\mcl P$, denote
by~$e_i$ the number of the flow edges, by~$n_i$ the number of
non-flow edges, by~$p_i$ number of flow edges pointing to a qubit to
be measured with a Pauli measurement and by $\ell_i$ the number of
loop edges ($\ell_i \in \{0,1\}$). Then the depth of the pattern,
call it~$D_{\mcl P}$ satisfies the following formula: \AR{ D_{\mcl
P} \leq \max_{I_i} e_i -(n_i+p_i + \ell_i) +1 \,.} \ORP

\begin{proof}
Along any Pauli influencing path, any flow edge pointing to a qubit
to be measured by a Pauli~$X$ will not require a separate layer
(Equation~\eqref{e-xrule}) and for the Pauli $Y$ case, such a flow
edge is converted to a~$Z$-dependency (Equation~\eqref{e-yrule}), to
be signal shifted as in Corollary~\ref{c-depth1}. Also if  the
influencing path starts with a~$Y$ measurement followed by a
non-Pauli measurement, we have a loop edge and hence the immediate
following non-Pauli measurement can also be put in the first layer
and hence we subtract the loop edge from the total depth for this
influencing path.
\end{proof}

\subsection{Classical depth}
\label{sec:classical depth} One issue that has often been overlooked
in the literature on MBQC is that computation of the correction
exponents as well as the measurement angles contributes to a
\emph{classical} depth~\cite{Jozsa05}. Consider, for example, the
case where we have a correction of the form $X_i^{s_1 + s_2 +\cdots+
s_n}$. An efficient implementation would start by classically
calculating the parity of the exponent, and then applying the
correction if the parity is $1$. This is also the case for a
measurement angle such as $\mR \al i {s_1 + s_2 +\cdots+ s_n}$,
where one needs to delay the quantum computation to classically
compute the measurement angle. Luckily all these classical delays
are of at most $O(\log(n))$ depth, since the parity of~$n$ bits can
be computed by a divide-and-conquer method in depth $O(\log(n))$
(any polynomial-size parity circuit has depth
in~$\Omega(\log^{*}\!n)$\cite{Parity84}). Such a classical
computation cost between quantum layers is negligible, but it still
exists. Actually, depending on the underlying  geometry of a pattern, this
classical processing sometimes requires  only constant depth. This
can be easily seen for a pattern with flow.

\LE\label{lem:classical measurement depth} Let $\mcl{P}$ be a
standard pattern  with flow and geometry $G$, before signal shifting
has been performed. The  depth of the classical processing required
between quantum layers is in~$O(\log\Delta(G))$. \EL
\begin{proof}
From the Flow Theorem, we know that each measurement at qubit~$i$
has at most one $X$-dependency from one of its neighbours in $G$ and
the rest of the neighbours of~$i$ contribute at most one
$Z$-dependency. Hence the depth for the classical computation
required for calculating the measurement angles at qubit~$i$ is
in~$O(\log(\deg(i)))$. Therefore, at each qubit, the classical depth
is in~$O(\log\Delta(G))$.
\end{proof}

Therefore, for a simple geometry  such as the cluster state with
maximum degree~4, all classical computation is constant. On the
other hand, signal shifting which is essential for decreasing the
quantum depth,  will increase the classical depth. We now present an
explicit quantum-classical tradeoff for  patterns with flow. But
first, we need to define a \emph{partial influencing path}: let~$v$
be a node in geometry~$G$. Then an~$I$ - $v$ partial influencing path
is a path in $G$ that starts with a flow edge at an input node
in~$I$, ends with a flow edge at node~$v$, and contains no
consecutive non-flow edges.

\begin{prop} \label{t-cdepth} Let~$\mcl P$ be a pattern with flow where
standardization and signal shifting have been performed. Fix a node
$v$ in the underlying geometry~$G$ and let~$I_v$ be the set of all
partial influencing paths, from an input qubit~$i$ to the node~$v$.
(If~$v$ is an output qubit we consider all the influencing paths
instead.) Let~$N_v$ be the set of vertices that are on any path
in~$I_v$. Then the classical depth of the required classical
computation for computing the angles of measurement command or the
exponent of correction command at~$v$ is in~$O(\log\abs{N_v})$.
\end{prop}

\begin{proof}
In the proof of Lemma~\ref{lem:classical measurement depth}, we saw
how the Flow Theorem tells us which dependencies are applied to a
qubit~$v$. Once signal shifting has been done, the dependencies are
modified, but they still propagate only through influencing paths.
In fact for the case of $v$ being a measured qubit, after signal
shifting only the $X$-dependencies remain and therefore we need to
consider only the partial influencing paths. Hence there are at
most~$\abs{N_v}$ dependencies at~$v$; the parity of these
dependencies can be computed in  classical depth~$O(\log\abs{N_v})$.
\end{proof}

\begin{figure}
\begin{center}
\includegraphics[scale=0.3]{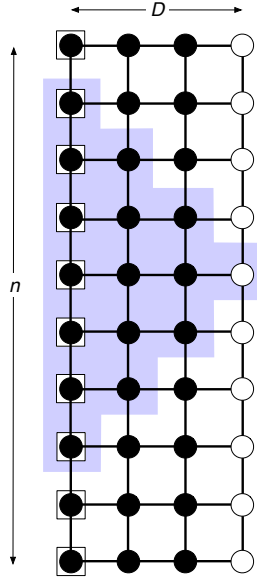}
\caption{A full cluster state geometry where the pyramid shape presents the backward cone, the set of all influencing paths that lead to a qubit. Only vertices on the pyramid contribute to the classical depth complexity of the command to be performed at that qubit.}
\label{pyramid}
\end{center}
\end{figure}

Note that $N_v$ is upper bounded by the total number of qubits in
the pattern. However for a particular geometry and angles of
measurement, it can be smaller. For example, consider a pattern with
$n$ input qubits and a geometry of a full cluster state of size $n$
times width equal to~$D$, as shown in Figure \ref{pyramid}. Then
from the above proposition, we conclude that the classical depth is in~$O(\log(D)$: for any given qubit $i$ only the $O(D^2)$
qubits siting on the pyramid with qubit $i$ as the top of the
pyramid, will contribute to the depth complexity of the command to
be performed at qubit $i$ (see Figure \ref{pyramid}). Therefore, for~$D\in O(\log(n))$, we obtain small classical depth of size
$O(\log(\log(n)))$, whereas the total number of the qubits in the
pattern is in~$O(n\log(n))$.

\section{Circuits and measurement patterns}\label{s-cir-pattern}

Having built all the required tools, we can now turn our attention to
the main focus of the paper on parallelizing quantum circuits. To
this end we give a method to translate a quantum circuit to a
pattern (Section~\ref{sec:circuits-to-patterns}) and vice-versa
(Section~\ref{sec:patterns-to-circuits}), where standardization,
signal shifting and Pauli simplifications on the obtained pattern
leads to a more parallel circuit. We also present the exact tradeoff
for the transformations. Furthermore, our construction allows us to
see influencing paths directly in the quantum circuit so that the
pattern depth characterization results given in
Section~\ref{s-charact} can be directly applied to circuits.

\noindent We  fix the universal family of  gates to be
$\mathfrak{U}=\{\ctR Z, J(\al)\}$: \AR{ \ctR Z= \MA{1&0&0&0\\
0&1&0&0\\ 0&0&1&0\\ 0&0&0&-1}, \,\,
J(\al)=\ost\MA{1&\ei\al\\1&-\ei\al}\,.}
 In
\cite{generator04,VC04} it was shown that this family is universal
for the circuit model since every single qubit unitary operator can
be written in terms of $J(\al)$: \AR{
U=e^{i\al}J(0)J(\ba)J(\ga)J(\da) } In addition, they lead to simple
generating patterns:
\begin{align}
\label{j-pattern} J(\al)&:=\cx 2{s_1}\m{{-\al}}1\et
12\\
\label{e-pattern} \ctR{Z}&:=\et 12 \,. \end{align} Hence this
family of unitaries is a good choice for translation between
circuits and patterns and any other universal family can be replaced
by this one with constant overhead.
In the rest of the paper, whenever the angle~$\al$ is not
important, we simply refer to a~$J(\al)$ gate as a~$J$ gate.

\subsection{From circuits to patterns}
\label{sec:circuits-to-patterns}

The original universality proof for MBQC already contained a method
to translate a quantum circuit containing arbitrary 1-qubit
rotations and control-not gates to a pattern~\cite{RB01}. Here, we
give an alternate method for the translation of a given circuit to a
standard pattern in the MBQC to attempt to reduce the quantum depth.
We give the exact tradeoff in terms of the number of auxiliary
qubits and depth.

Recall that $\ctR Z$ is self-inverse and symmetric, hence any circuit that
contains consecutive~$\ctR Z$ gates acting on the same qubits can be
simplified. In what follows, we suppose that this simplification has
been performed.

\DE \label{def:Circuit-to-pattern} Let $C$ be a circuit of $\ctR Z$ and $J$
gates on $n$ logical qubits. The corresponding standard pattern $\mathcal{P}$ is
obtained by replacing each gate in $C$ with its corresponding
pattern given by equations \eqref{j-pattern} and \eqref{e-pattern},
and then performing standardization and signal shifting. \ED

To present the exact tradeoff for the above translation, in
particular to prove that the quantum depth cannot increase, we
construct directly the underlying geometry of a given circuit.
Following the literature, we refer to the circuit qubits as
\emph{logical} qubits. Other qubits that are added during
construction of the entanglement graph will  be referred to as
\emph{auxiliary} qubits.

\begin{figure}
\begin{center}
\includegraphics[scale=0.65]{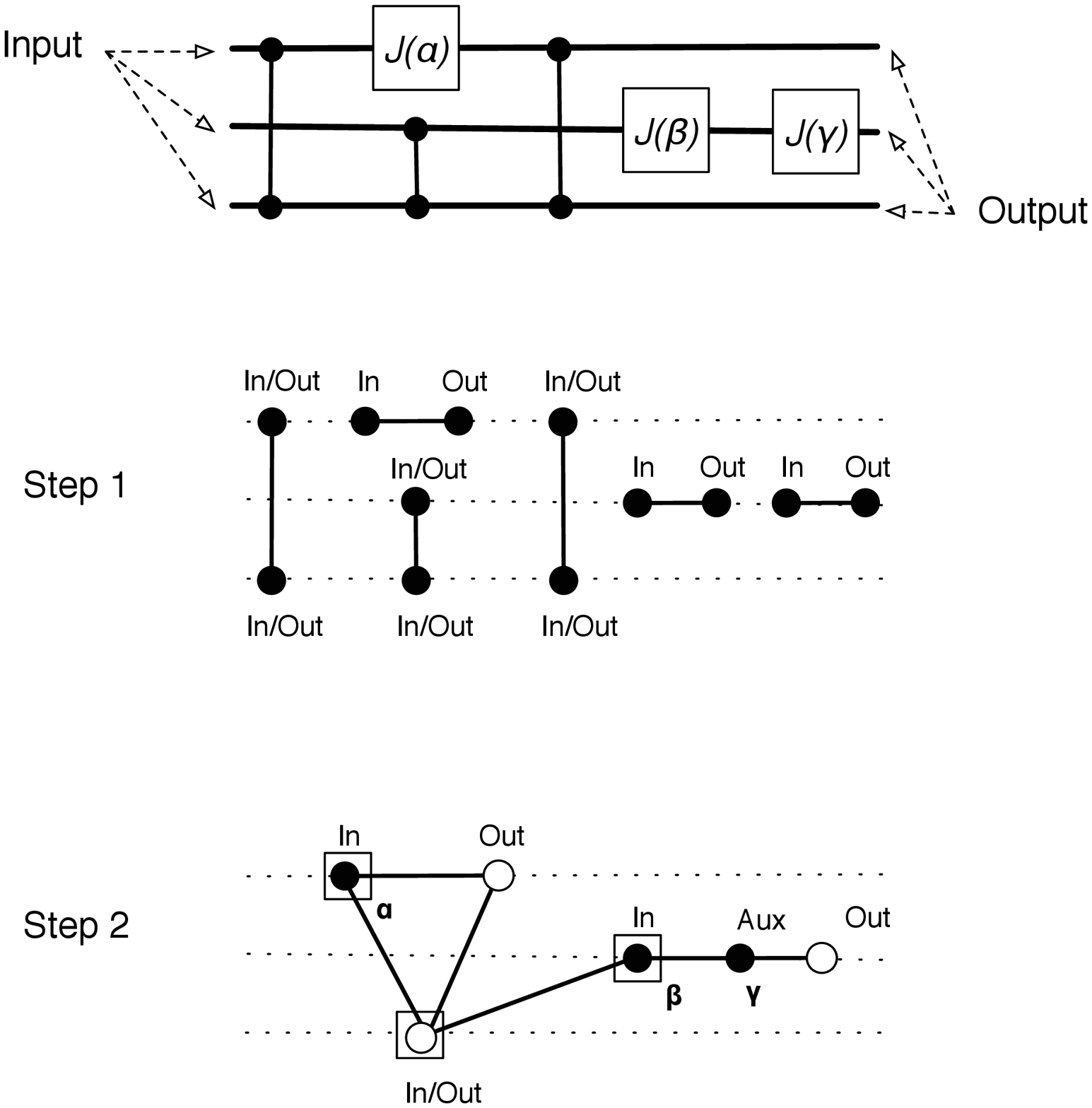}
\caption{A quantum circuit with $\ctR Z$ and $J(\al)$ gates,
together with the two-step construction of the corresponding
labelled entanglement graph. In the final step, an input qubit is
represented by a boxed vertex and an output qubit with a white
vertex. The black vertices will be measured with angles $\al,\ba$
and $\ga$, as shown in the figure.} \label{circgraph}
\end{center}
\end{figure}

\DE\label{d-cgraph} Let $C$ be a circuit of $\ctR Z$ and $J$
gates on $n$ logical qubits. The \emph{labelled entanglement graph} $G_C$ is
constructed  as a layer that is initially built on top of the
circuit~$C$ by the following steps (see also the example of
Figure~\ref{circgraph}).
\begin{itemize}

\item[1.]
Replace each $\ctR Z$ gate on logical qubits~$i$ and~$j$ with a
vertical edge between two vertices: one on the~$i^{\text{th}}$ wire
and one on the~$j^{\text{th}}$ wire. Label both vertices
\emph{Input/Output}. Replace each~$J$ gate on a logical
qubit~$i$ with an horizontal  edge between two vertices on
the~$i^{\text{th}}$ wire, label the left vertex \emph{Input} and the
right vertex \emph{Output}.

\item[2.] To connect the above components,
on each wire, start from the left  and contract consecutive
non-adjacent vertices as follows (the \emph{contraction} of
vertices~$v_1$ and~$v_2$ of a graph $G$ is obtained by
replacing~$v_1$ and~$v_2$ by a single vertex~$v$, which is  adjacent
to all the former neighbours of~$v_1$ and~$v_2$):

\begin{itemize}
\item Two vertices labelled \emph{Input/Output} are contracted as one vertex with \emph{Input/Output} label;
\item A vertex labelled  \emph{Input/Output} and a vertex labelled \emph{Input} are contracted
as one vertex with \emph{Input} label;
\item A vertex labelled  \emph{Output} and a vertex labelled  \emph{Input/Output}
 are contracted as one vertex with \emph{Output} label;
\item Two vertices labelled  \emph{Output} and \emph{Input} are contracted as one vertex with \emph{auxiliary}
label.
\end{itemize}
\end{itemize}
\ED

It is easy to verify the following proposition that justifies the
above construction.

\PRO\label{p-cgraph} The graph $G_C$ obtained from
Definition~\ref{d-cgraph} is the entanglement graph for the
measurement pattern that is obtained from
Definition~\ref{def:Circuit-to-pattern}.  Furthermore, input-output
paths of vertices sitting on the same wire define the flow of~$G_C$.
\ORP
\begin{proof}
Standardization does not change the underlying entanglement
graph, hence it follows that~$G_C$ is indeed the entanglement graph for
the measurement pattern. By Theorem~10 of~\cite{Beaudrap06}, for the
case that~$\abs{I} = \abs{O}$, a collection of vertex-disjoint $I-O$
paths in~$G_C$ define its flows. Therefore, input-output paths of
vertices sitting on the same wire define the flow of~$G_C$.
\end{proof}

In order to obtain a full pattern corresponding to the circuit~$C$,
one needs to add measurement commands with angles being the same
angles of the~$J(\al)$ gates. These angles are assigned to the
qubits labelled \emph{Input} in Step~(1) of the construction of
Definition \ref{d-cgraph}. The dependency structure is the one
obtained from the Flow Theorem.

\PRO \label{p-cir2pat} Let $C$ be a quantum circuit on~$n$ logical qubits
with only~$\ctR Z$ and~$J$ gates. Let~$G_2$ be the number of
$J$ gates and~$D(n)$ the circuit depth. The corresponding
pattern $\mathcal{P}$ given by Definition~\ref{def:Circuit-to-pattern} has~$n+G_2$ qubits,~$G_2$ measurement
commands, $n$~corrections commands, and depth smaller than or equal
to~$D(n)$. \ORP

\begin{proof}
The proof is based on construction of Definition~\ref{d-cgraph},
which is obtained from replacing the patterns from
equations~\eqref{j-pattern} and~\eqref{e-pattern} for~$J$
and~$\ctR Z$ gates and then performing the standardization
procedure. It is clear from the construction that we start with $n$
qubits corresponding to each wire, then any~$\ctR Z$ connects the
existing qubits (wires) and hence will not add to the total number
of qubits. On the other hand any $J$ gate extends the wire by
adding a new qubit. This  leads to the total number of~$n+G_2$
qubits for the pattern. There are~$G_2$ measurement commands since
all but~$n$ qubits are measured.
Since~$C$ has depth~$D(n)$, any influencing path in~$\mathcal{P}$
has at most~$D(n)$ flow edges.
 Hence the theorem is obtained from Proposition~\ref{c-depth1}
after performing signal shifting on the corresponding pattern.
\end{proof}

\begin{figure}
\begin{center}
\includegraphics[scale=0.4]{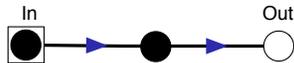}
\caption{The geometry of the teleportation pattern given in Equation \eqref{e-teleport} with one input, one auxiliary and one output qubit. }
\label{teleport}
\end{center}
\end{figure}

Alternatively, for a given circuit, one can use another construction
to obtain a corresponding pattern with cluster geometry, hence to
achieve constant depth for the graph preparation stage. Naturally,
the price is to have more qubits. First note that the following
pattern implements teleportation from input qubit~$i
$ to output qubit~$k$ that is simply the identity map~(see Figure \ref{teleport}):
\EQ{\label{e-teleport} \cx k {s_j}\cz k {s_i} \m 0 j \m 0 i \et jk \et ij }

Now, if before Step~(2) of the construction of Definition
\ref{d-cgraph}, we insert the teleportation pattern  between any two
consecutive $\ctR Z$ acting on a common wire, then the degree of
each vertex remains less than~4 as desired. We will refer to this
graph as the \emph{cluster graph}, ${GC}_C$. In order to compute the
number of qubits for the pattern obtained from this new
construction, consider the positions in the circuit
where two~$\ctR Z$ appear after each other. These are the places
where we need to apply the above teleportation pattern to keep the
degree less than~4. With this construction, the depth of the pattern
does not increase by more than a multiplicative constant. Therefore
we have:

\LE \label{l-cluster} Let $C$ be a quantum circuit on~$n$ qubits
with only~$\ctR Z$ and~$J$ gates. Let $G_2$ be the number of
$J$ gates, $s$ the size of~$C$ and $m$ the number of positions
in~$C$ where two~$\ctR Z$ appear after each other. Then the pattern
$\mathcal{P}$ with the cluster graph construction (obtained as in
Proposition~\ref{p-cgraph} with the addition of the teleportation
pattern above)
 has
\mbox{$n+G_2+m \in O(n+s)$} qubits and depth in~$O(D(n))$. \EL

In what follows, we always assume the cluster geometry for patterns
corresponding to a circuit and hence the preparation depth is~$4$
(Section~\ref{sec:preparation depth complexity}).

\subsection{From patterns to circuits}
\label{sec:patterns-to-circuits}

\begin{figure}
\begin{center}
\includegraphics[scale=0.5]{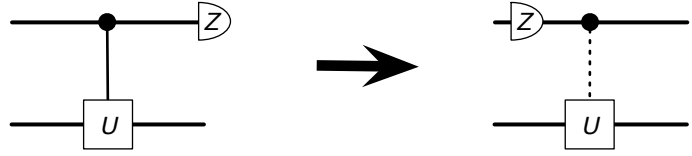}
\caption{A classically controlled implementation of a
controlled-unitary gate.  The computational basis measurement
operator is represented by the half-circle box with $Z$ label. After
pushing the measurement to the beginning of the wire, the
unitary~$U$ is only classically dependent (doted line) on the first
wire.} \label{f-ctrl-U}
\end{center}
\end{figure}

The construction of Definition \ref{d-cgraph} can be also used in
reverse order to transfer a pattern with flow to a corresponding
circuit, where all the auxiliary qubits will be removed and hence by
doing so the quantum depth might increase. However, we now show  how
to obtain another transformation from patterns to circuits where one
keeps all the auxiliary qubits. This new construction is simply
based on the well-known method of coherently implementing a
measurement. Recall that a controlled-unitary operator where the
control qubit is measured in the computational basis~\{$\ket 0,\ket
1$\} can be written as a classical controlled unitary by pushing the
measurement before the controlled-unitary operator~\cite{GN96}, see
Figure~\ref{f-ctrl-U}.

\begin{figure}
\begin{center}
\includegraphics[scale=0.37]{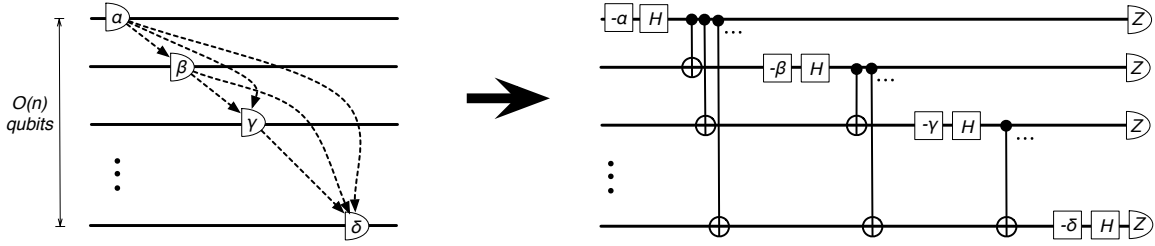}
\caption{Implementing coherently the sequence of dependent measurements in a pattern. An arbitrary measurement~$M^\al$ is represented by a half circle labelled with its angle. The Hadamard and phase gates are shown with square boxes with the labels being~$H$ or the angle of the phase gate. The dotted arcs represent $X$-dependencies. Equation~\eqref{e-simp} is used to simplify the measurements. After replacing the $X$-dependencies by $\ctR X$ gates, we obtain a quantum depth linear in the number of dependencies.}
\label{f-C-control}
\end{center}
\end{figure}

Given a pattern in the standard form, we use the above scheme in the
reverse order to convert the classically dependent measurements and
corrections, and then push all the independent measurements to the
end of the pattern. However since the scheme works only for the
computational basis measurement, we have to first simplify all the
arbitrary measurements~$M^\al$. Let~$Z(\al)$ be the phase gate
and~$H$ the Hadamard gate (see Appendix \ref{Appendix-B}), and
let~$M^Z$ be the computational basis measurement (\ie~Pauli $Z$
measurement). Then we have \EQ{ \label{e-simp} M^{\alpha} =
M^{\{\ket {+_\al},\ket {-_\al}\}}=M^{HZ(-\al)^{\dag}\{\ket 0,\ket
1\}}=M^Z H Z(-\al)\,. } Additionally, we replace any classical $X$-
and $Z$-dependencies of measurements and any dependent corrections
with a sequence of~$\ctR X$ and~$\ctR Z$, which might creates a
quantum depth linear in the number of the dependencies, as shown in
Figure~\ref{f-C-control}. However to reduce this linear depth, we
can use the following result on parallelizing a circuit with only
controlled-Pauli gates to logarithmic depth:

\PRO {(\cite{MN02})} \label{p-moore} Circuits on~$n$ qubits consisting
of controlled-Pauli gates and the Hadamard gate can be parallelized
to a circuit with~$O(\log n)$ depth and~$O(n^2)$ auxiliary qubits.
\ORP

We can now formalize the above translation of patterns to circuits.

\DE\label{def:patterntocircuit} Let~$\mcl P$ be a standard pattern
with computational space~$(V,I,O)$, underlying geometry~$(G,I,O)$
(where~$G$ has a constant maximum degree) and command sequence
(after signal shifting): \AR{
 \cdots C_j^{C_j} \cdots \mR {\al_i}i{A_i} \cdots E_G
} where~$A_i$ is the set of qubits that the measurement  of
qubit~$i$  depends on, and~$C_j$ is the set of qubits that the
correction of qubit~$j$ depends on.  Note that due to the signal
shifting, we only have~$X$ dependencies. The corresponding
\emph{coherent circuit}~$C$ with~$|I|$ logical qubits
and~$|V\setminus I|$ auxiliary qubits, is constructed in the
following steps (see also Figure~\ref{f-C-control}):
\begin{enumerate}
\item \label{p-c:1} Apply individual Hadamard gates on all the auxiliary qubits.
\item \label{p-c:2} Apply a sequence of $\ctR Z$ gates according to the edges of $G$.
\item  \label{p-c:3} Replace any dependent measurement $\mR {\al_i}i{A_i}$ with~$M_i^Z H_i Z_i(-\al)\ctR_{A_i,i} X$
where $\ctR_{A,i} X$ is a sequence of controlled-not with control qubits in~$A$ and target qubit~$i$. Note that
since the~$M^Z$ is independent and can be pushed to the end of the corresponding wire it can be discarded.
\item\label{p-c:4}  Replace any dependent correction~$X_j^{C_j}$ with~$\ctR_{C_i,i} X$ and~$Z_j^{C_j}$ with~$\ctR_{C_i,i} Z$.
\item \label{p-c:5} Replace the joint sequence of added~$\ctR X$ and~$\ctR Z$ in steps~\ref{p-c:3} and~\ref{p-c:4} with the
parallel form obtained from Proposition~\ref{p-moore}.
\end{enumerate}
\ED

\begin{lemma}
\label{lem:cost-patterntocircuit} Let~$\mcl P$ be a standard pattern
with computational space~$(V,I,O)$ and underlying geometry~$(G,I,O)$
(where~$G$ has a constant maximum degree). Let~$t=\abs{V \setminus
O}$ be the number of measured qubits and let~$d$ be the quantum
computation depth of~$\mcl P$. Then the corresponding coherent
circuit~$C$ obtained from Definition~\ref{def:patterntocircuit}
has~$\abs{I}$ logical qubits,~$O(t^3)$ auxiliary qubits and depth
in~$O(d\log t)$.
\end{lemma}
\begin{proof}
We examine the cost at each step of the construction of
Definition~\ref{def:patterntocircuit}. Steps~\ref{p-c:1}
and~\ref{p-c:2} add a constant to the depth of~$C$. At
step~\ref{p-c:3}, each measurement has as most~$t$ dependencies,
which, in step~\ref{p-c:5} translates to~$O(\log t)$ depth
with~$O(t^2)$ auxiliary qubits. At step~\ref{p-c:4}, each output
qubit has at most~$t$ dependencies, which again in step~\ref{p-c:5}
translates to~$O(\log t)$ depth with~$O(t^2)$ auxiliary qubits.
Since the depth of~$\mcl P$ is~$d$, the total depth of~$C$ is
in~$O(d\log t)$, with $O(t^3)$ auxiliary qubits.
\end{proof}

Note that the logarithmic increase in the depth of~$C$ is due to the
fact that the circuit model does not exploit any classical
dependencies. Thus the classical computation of the measurement
angles and corrections in~$\mcl P$ contributes to the quantum
depth in~$C$.

One can combine the forward and backward construction from circuit
to patterns to obtain an automated rewriting system for the circuit
which can decrease the depth by adding auxiliary qubits. The
following theorem gives the tradeoff.

\TH \label{t-cir2pat2cir} Let $C$ be a quantum circuit on $n$ qubits
with only~$\ctR Z$ and~$J$  gates. Suppose~$C$ has size~$s$ and
depth~$D$. Assume further that~$\mathcal{P}$ is the corresponding
pattern obtained from the forward translation as in
Lemma~\ref{l-cluster} and that~$\mathcal{P}$ has quantum depth~$D'$
(we know that~$D' \leq D$).  Then circuit~$C'$ constructed
from~$\mathcal{P}$ by Definition~\ref{def:patterntocircuit}
has~$O(s^3+n)$ qubits, and depth in~$O(D' \log s)$. \HT

\begin{proof}
The first step is to translate~$C$ to a pattern~$\mathcal{P}$ using
Lemma~\ref{l-cluster}. The resulting pattern~$\mathcal{P}$
has~$O(s+n)$ qubits, and quantum depth in~$O(D)$. Then we translate
the pattern back to a circuit~$C'$ using
Definition~\ref{def:patterntocircuit}. By
Lemma~\ref{lem:cost-patterntocircuit}, the new circuit has $O(s^3)$
auxiliary qubits and depth in~$O(D'\log s)$.
\end{proof}

At first glance it seems like applying Theorem~\ref{t-cir2pat2cir}
to a quantum circuit would not necessary be beneficial, since the
number of auxiliary qubits and the depth seem to increase. But note
that we have given only upper bounds. As we showed in
Section~\ref{s-flowdepth}, taking into account Pauli simplification
and signal shifting can give a significant improvement. In the next
section, we give a complete depth characterization for patterns with
flow, and then show in Section~\ref{sec:parallizing circuits} a
characterization of those circuits to which applying Theorem
\ref{t-cir2pat2cir} will necessarily decrease the depth.

\section{Depth Characterization}\label{s-charact}

We saw in Section~\ref{s-flowdepth} that the main ingredients to
obtain a reduced pattern depth are influencing paths, Pauli
measurements and signal shifting. In fact, Pauli measurements not
only can be performed in the first layer but also they can ``reset''
the pattern depth along an influencing path. This intuition is
formalized in the following lemmas that are essential for our
characterization result. In what follows, we deal with sequences
of measurements angles, where~$N_i, N_j, \dots$ represent non-Pauli
measurements, $X$ a Pauli $X$ measurement, $Y$ a Pauli $Y$ measurement and~$P$ is either~$X$ or~$Y$.
Furthermore $(\omega)^*$ and $(\omega)^{\text{odd}}$ represents respectively, a
non-negative and odd number of repetitions of~$\omega$.

\LE \label{l-nodepend} Let $N_i$ and $N_j$ denote two non-Pauli
measurements on a common influencing path~$I$ of a standard pattern with
flow. Suppose that~$N_i$ and~$N_j$ are separated along~$I$ with only
flow edges, and that the sequence of measurements between~$N_i$
and~$N_j$ along~$I$ is a sequence of Pauli measurements of the form:
 \AR{
 (X)^{\text{odd}}(Y(X)^{\text{odd}})^*\, .
 }
Then after signal shifting, there will be no $X$-dependency between $N_i$ and~$N_j$. \EL

\begin{figure}
\begin{center}
\includegraphics[scale=0.5]{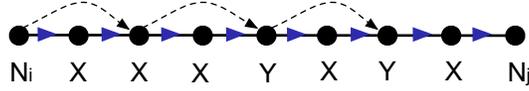}
\caption{Two non-Pauli measurements separated with the sequence of Pauli measurements of the form $(X)^{\text{odd}}(Y(X)^{\text{odd}})^*$. There is no $Z$-dependency between the last Pauli $X$ measurement and the first non-Pauli measurement and therefore, after signal shifting, there will be no $X$-dependency between the non-Pauli measurements.}
\label{evenY}
\end{center}
\end{figure}

\begin{proof}
Assume such an $X$-dependency between $N_i$ and~$N_j$ exists, then
it is necessarily due to the fact that during signal shifting, the
last Pauli $X$ measurement in the sequence acquires a $Z$-dependency
from~$N_i$; this $Z$-dependency would then be signal shifted to an
$X$-dependency between $N_i$ and~$N_j$, since $N_j$ has an
$X$-dependency on the last Pauli $X$ measurement. We use a parity
argument to show that this never occurs.

First, note that the sequence of Pauli measurements,
$(X)^{\text{odd}}(Y(X)^{\text{odd}})^*$ is odd. Second, note that
through signal shifting, the $Z$-dependency that originates
from~$N_i$ is shifted only through every even position in the Pauli
measurement sequence. Due to the placement of the $Y$
measurements which never occur in an odd position, the
special case of the Pauli $Y$ rule (Equation \eqref{e-yrule}) cannot
be applied to change the parity. Hence, the final~$X$ measurement in
the sequence (which is at an odd position) never sees a $Z$-dependency
from~$N_i$ (Figure \ref{evenY}).
\end{proof}

\LE \label{l-indirectX} Let $N_i$ and $N_j$ denote two non-Pauli
measurements on a common influencing path~$I$ of a standard pattern with
flow.  Suppose that~$N_i$ and~$N_j$ are separated along~$I$ with
 only Pauli measurements, \ie~we have the following sequence along~$I$:
\AR{ N_i \; \msf P_1\; \al_1\ba_1 \; \msf P_2\; \al_2\ba_2 \; \cdots
\; \msf P_k\; N_j \, , } where $\msf P_i$ is a (possible empty)
finite sequence of Pauli measurements and $\al_i\ba_i$ represents
the endpoints of a non-flow edge. After signal shifting, there will
be no $X$-dependency due to~$I$ between $N_i$ and~$N_j$ if and only
if at least one of the $\msf P_i$ sequence is equal
to~$(X)^{\text{odd}}(Y(X)^{\text{odd}})^*$. This will be also true
even if one of the $N_i$ or $N_j$ are an endpoint of a non-flow
edge.\EL

\begin{proof} First, assume $N_i$ and $N_j$ are connected with only flow
edges (we have the sequence~$N_i \msf P N_j$) and consider the following
possible cases for the sequence~$\msf P$:
\begin{enumerate}
\item[(I).]  It consists of an even number of Pauli angles. Then
there is an $X$-dependency between~$N_i$ and~$N_j$.
\item[(II).] It consists of an odd number of  Pauli angles with at least one~$Y$ at an odd position from left to
right. Then there is an $X$-dependency between~$N_i$ and~$N_j$.
\item[(III).] It consists of an odd number of  Pauli angles with no~$Y$ at any odd
position: $(X)^{\text{odd}}(Y(X)^{\text{odd}})^*$. Then there is
\emph{no}  $X$-dependency between~$N_i$ and~$N_j$.
\end{enumerate}

\begin{figure}
\begin{center}
\includegraphics[scale=0.5]{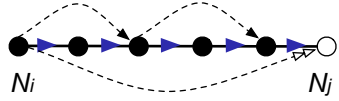}
\caption{An even number of Pauli measurements between two non-Pauli
measurement leads to an $X$-dependency after signal shifting.}
\label{f-evenP}
\end{center}
\end{figure}

Figure \ref{f-evenP} shows how in Case~(I), one obtains an
$X$-dependency after signal shifting between $N_i$ and $N_j$.
Case~(II) is also similar, by Pauli simplification, the
$X$-dependency at a $Y$ measurement of an odd position is considered
as an $Z$-dependency  and hence we obtain the same scenario as
Case~(I). Finally,  Case~(III) is proved in Lemma \ref{l-nodepend}.

Now consider the case where there exists a non-flow edge between the
non-Pauli angles and neither~$N_i$ nor~$N_j$ are an endpoint of a non-flow edge:
\AR{ N_i \; \msf P_1\; \al\ba \; \msf P_2 N_j \, . }
According to the Flow Theorem, there is a $Z$-dependency from the
qubit that precedes the  qubit assigned to~$\al$ angle to the qubit
with angle~$\ba$. In order to have a sequence of $Z$ dependencies
between $N_i$ and~$\ba$, $\msf P_1$ must satisfy the conditions of
cases~(I) or~(II) and then similar to the above argument, in order
to obtain an $X$-dependency between $N_i$ and~$N_j$, $\msf P_2$ must also
satisfy the conditions of cases~(I) or~(II) and hence we obtain the
statement of the Lemma. The same argument is valid if either of $N_i$ or
$N_j$ is an endpoint of a non-flow edge.
\end{proof}

We can now present our main result on characterization of patterns with a given depth.

\begin{theorem} \label{t-depthd} Let~$\mcl P$ be a standard pattern  with flow and let~$I$ be an
influencing path of~$\mcl P$. We apply the following simplification rule to the
commands along~$I$:
\begin{equation*}
 N \, \msf P_1\, \al_1\ba_1 \, \msf P_2\, \al_2\ba_2 \,
\cdots \, \msf P_k\, N \Rightarrow
\begin{cases}
N \hskip 1cm \text{if} \,\,\, \exists \msf P_i=(X)^{\text{odd}}(Y(X)^{\text{odd}})^* \\
NN \hskip 0.7cm \text{otherwise}\,.
\end{cases}
\end{equation*}
where $\msf P_i$ represents a (possible empty) finite sequence of
Pauli measurements and $\al_i\ba_i$ represents the endpoints of a
non-flow edge. Define the depth of~$I$ to be~$d+2$ if after the
simplification we obtain $\msf P\,N^{d}\,\msf P$ and to be $d+1$ if
we obtain either $Y\,N^{d}\,\msf P$ or $N^{d}\,\msf P$. Then the
quantum depth of~$\mcl P$ after Pauli simplification and signal
shifting is given by the maximum depth over all influencing paths
of~$\mcl P$.
\end{theorem}

\begin{proof}
Lemma \ref{l-indirectX} justifies the given simplification rule. It
is trivial that after applying the rule one will obtain a unique
final sequence of the form $\msf P\,N^{i}\,\msf P$ on any influencing path and
hence the longest sequence of dependent of non-Pauli measurements
will have length~$i$ and since there is a first layer of Pauli and
one final layer of corrections the depth along this path will be
$i+2$. However if the final form is $Y\, N^i \,\msf P$ then there will be
no dependency between the Pauli~$Y$ and the first non-Pauli~$N$
(Equation \eqref{e-yrule}) and depth is $i+1$ which is also the case
for the final form $N^i\,\msf P$.

According to  Proposition \ref{prop:depend-ipaths} the pattern depth
is the maximum number of the dependent non-Pauli measurements along
all the influencing paths and hence it is enough to compute the
maximum value of~$i$  over all influencing paths. \end{proof}

The above theorem gives a constructive method to obtain a depth $d$
pattern. The main tool being the sequence
$(X)^{\text{odd}}(Y(X)^{\text{odd}})^*$, which if it is inserted
between two non-Pauli measurements make them independent of each
other. On the other hand,
any other sequence inserted between non-Pauli angles
contributes  to the depth and makes the two non-Pauli measurements
$X$-dependent on each other and hence in two different layers of
measurement.

We now show as a special case the characterization of patterns with
depth~2.

\PRO \label{p-depth1} Let $\mcl P$ be a pattern with flow~$f$, where
standardization, Pauli simplification and signal shifting have been
performed. The quantum computation depth is equal to~2 if and only
if any qubit measured with a non-Pauli angle is not the flow image
of any other vertex and hence it is either an input qubit or is
connected to a vertex with a loop flow edge. \ORP

\begin{proof} Due to Theorem~\ref{t-depthd}, $\mcl P$  has depth~2
if and only if on all the influencing paths, after the
simplification rule, we obtain one of the following final forms for
the sequence of the measurement angles: \AR{ N\, \msf P \;\;\;
\text{or} \;\;\; Y\,N\, \msf P \;\;\; \text{or} \;\;\; \msf P\,. }
Now consider  only those influencing paths with only flow edges, by
reverse application of the simplification rules we conclude only
input qubits can be measured with a non-Pauli angle or a non-input
qubit measured by a non-Pauli measurement should not be the flow
image of any other qubit and be connected to a qubit measured with
Pauli~$Y$.
\end{proof}

Note that this proposition extends the previously know result that
patterns with only Pauli measurements have
depth~2~\cite{Jozsa05,RBB03}.

\subsection{Parallelizing Circuits}
\label{sec:parallizing circuits}

\begin{figure}
\begin{center}
\includegraphics[scale=0.5]{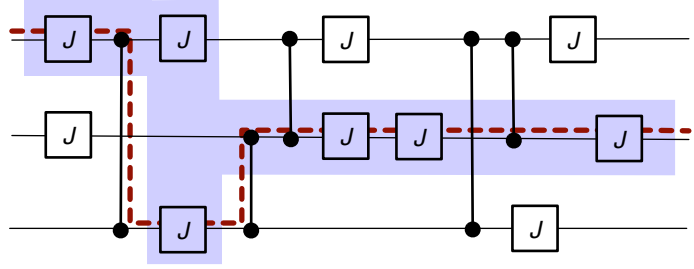}
\caption{A circuit with one of its influencing path presented as a
doted line. The $J$ gates in the shaded area are those referred to
as the $J$ gates of the path.} \label{cinflupath}
\end{center}
\end{figure}

In order to present the pattern depth characterization result
directly in terms of the circuit language, we first define the notion
of \emph{circuit influencing paths}.

\DE Let $C$ be a circuit of $\ctR Z$ and $J$ gates. A left-to-right path starting
at the beginning of a circuit wire and ending at any wire, such that the jumps between wires are done
through~$\ctR Z$ gates is called a \emph{circuit influencing path}
if there exist no two consecutive jumps (see Figure~\ref{cinflupath}). \ED

Recall that for patterns with equal number of input and output qubits, the
flow, if it exists, is unique. Hence it is easy to verify that
circuit influencing paths defined above are exactly influencing
paths of the corresponding pattern via the direct translation given in
Section~\ref{s-cir-pattern}. Similar to the pattern case, the circuit
depth is characterized in terms of the sequence of~$J$
gates appearing on the influencing paths defined below.

\DE \label{d-Jgates} Let $I$ be a circuit influencing path of
circuit~$C$. The set of $J$ gates over $I$ is defined to be all the
consecutive $J$ gates over the wires of the path including the $J$
gates just after a $\ctR Z$ gate of a jump, as shown in Figure
\ref{cinflupath}. \ED

Note that, again the above definition is a direct consequence of our
transformation between circuits and patterns. Further, define $H^i$ to
be the single unitary gate \AR{ \frac{1}{\sqrt{2}}\begin{pmatrix} 1
& -i \\ 1 & i\end{pmatrix} } implemented by the pattern $\cx
2{s_1}\m{{\pit}}1\et12$. We also have, $J(0)=H$ and $J(\pit)=H^i$.
We can now present our depth result directly for circuits.

\TH \label{t-cdepthd} Let $C$ be a circuit of $\ctR Z$ and $J$ gates
on $n$ qubits with size~$s$ and depth~$D$. Assume that after the
following simplification rule on $J$ gates over all circuit
influencing paths, we obtain at most $D'$ many consecutive $J$ gates:
\begin{equation*}
 J \, \msf P_1\, \al_1\ba_1 \, \msf P_2\, \al_2\ba_2 \,
\cdots \, \msf P_k\, J \Rightarrow
\begin{cases}
J \hskip 1cm \text{if} \,\,\, \exists \msf P_i=(X)^{\text{odd}}(Y(X)^{\text{odd}})^* \\
JJ \hskip 0.7cm \text{otherwise}\,.
\end{cases}
\end{equation*}
where $\msf P_i$ represents a (possible empty) finite sequence
of~$H$ and~$H^i$ gates and~$\al_i\ba_i$ represents the~$J$ gates
immediately after a~$\ctR Z$ gate on the underlying circuit
influencing path. Then, using the construction of Section
\ref{s-cir-pattern}, circuit~$C$ can be parallelized to an
equivalent circuit~$C'$ with depth in~$O(D'\log(s))$ and size
in~$O(s^3+n)$. \HT

\begin{proof}
The proof simply follows from theorems~\ref{t-cir2pat2cir}
and~\ref{t-depthd}.
\end{proof}

Similar to the pattern case, the above theorem gives a constructive
method to obtain a depth~$d$ circuit. The main tool is the gate
sequence \EQ{\label{e-reset}
R=(H)^{\text{odd}}(H^i(H)^{\text{odd}})^*\, ,} which if it is
inserted between two $J$ gates over a circuit influencing path will
make them to appear in the same layer of the final parallelized
circuit.

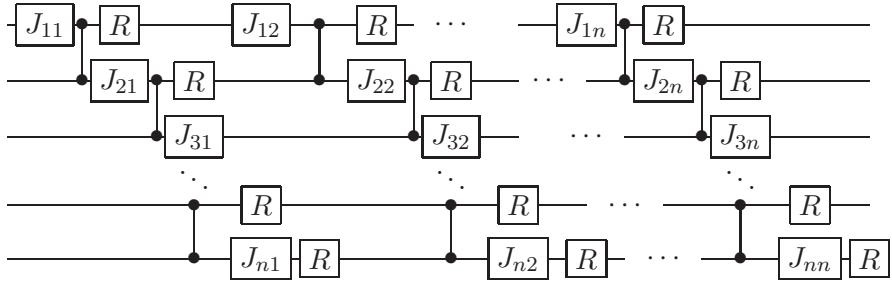
\begin{figure}[h!]
\[
\Qcircuit @C=.3em @R=.5em {
\hspace{-.7cm}
&&\gate{J_{11}} & \ctrl{1} & \gate{R} &  \qw& \qw &\gate{J_{12}} & \ctrl{1} & \gate{R} &  \qw &\cdots & &\gate{J_{1n}} & \ctrl{1} & \gate{R}  & \qw & \qw & \qw & \qw\\
\hspace{-.7cm}
&& \qw & \ctrl{-1}&  \gate{J_{21}} & \ctrl{1} & \gate{R} & \qw& \ctrl{-1}&  \gate{J_{22}} & \ctrl{1} & \gate{R} &  \qw & \hspace{-1cm} \cdots   &  \ctrl{-1}&  \gate{J_{2n}} & \ctrl{1} & \gate{R} & \qw & \qw \\
\hspace{-.7cm}
&& \qw & \qw &\qw & \ctrl{-1} &  \gate{J_{31}} &  \qw & \qw & \qw&  \ctrl{-1} &  \gate{J_{32}} & \qw & \cdots & &  \qw&  \ctrl{-1} &  \gate{J_{3n}}& \qw   &\qw \\
\hspace{-.7cm} \vspace{1cm}  & \vspace{10cm} &  & & &  &\ddots&  & & &   &\ddots & &  & & & &\ddots \\
\hspace{-.7cm}&&   \qw& \qw& \qw & \qw  &  \ctrl{1}& \gate{R} & \qw& \qw  & \qw  &   \ctrl{1} & \gate{R}&  \qw &\cdots& & \qw  &  \ctrl{1} & \gate{R} & \qw \\
\hspace{-.7cm}&& \qw  &  \qw& \qw  & \qw  & \ctrl{-1} &
\gate{J_{n1}} & \gate{R}& \qw&  \qw  & \ctrl{-1} &
\gate{J_{n2}} &  \gate{R}& \qw & \cdots & & \ctrl{-1} &
\gate{J_{nn}} & \gate{R} }
\]
\caption{A polynomial-depth circuit where each $J_{ij}$ gate has
an angle~$\in[0,2\pi)$ and the~$R$ gate stands for a
sequence of Clifford gates of the form
$(H)^{\text{odd}}(H^i(H)^{\text{odd}})^*$. Theorem \ref{t-cdepthd}
implies that this circuit can be parallelized to a logarithmic depth
circuit.} \label{polydepth}
\end{figure}

As an application, consider the quantum circuit  in Figure~\ref{polydepth} with size in~$O(n^2)$ and depth in~$O(n)$. Theorem
\ref{t-cdepthd} tell us how to parallelize it to depth
in~$O(\log(n))$, while adding~$O(n^6)$ auxiliary qubits. First
note that on any circuit influencing path, any two~$J$ gates are
separated by an~$R$ gate (Equation \eqref{e-reset}) and hence
after the simplification rule, we will have no two consecutive $J$ gates.
In other words, the parameter $D'$ in Theorem \ref{t-cdepthd} is equal
to 1 which implies the depth of the parallelized circuit will be in~$O(\log(n))$.

It is easy to extend the circuit of Figure \ref{polydepth} and
still apply Theorem \ref {t-cdepthd} to parallelize it to a circuit
with depth in $O(\poly(\log(n)))$. On each wire, replace
$O(\log(n))$ many $J_{ij}$ gates with the following sequence of
gates: \AR{ J_1 \msf P_1 J_2 \msf P_2 \dots J_{k} \hspace{0.5cm}
\text{with} \hspace{0.5cm} k\in O(\log (n)) } where $\msf P_i$ is a sequence of~$H$ and~$H^i$ gates of polynomial length.
Now the parameter
$D'$ of Theorem \ref{t-cdepthd} is in~$O(\log(n))$ and the parallel
circuit will have depth in~$O(\log^2(n))$.

These set of examples, although somewhat artificially constructed,
demonstrate how one might use Theorem~\ref{t-cdepthd} to construct
parallel circuit for a given problem in hand. We finish this section
with several other results on circuit parallelization.

\PRO\label{p-cdepth2} A circuit on $n$ qubits can be parallelized to
a pattern of depth~2 via the construction given in
Section~\ref{s-cir-pattern} if and only if it is   of the form: a
possible sequence of individual phase gates, $Z_1(\al_1) \otimes
\cdots \otimes Z_n(\al_n)$, followed by an arbitrary poly-size
Clifford circuit.
\ORP

\begin{proof}
It is known that any Clifford gate can be implemented by a pattern
with only Pauli~$X$ and~$Y$ measurements \cite{DKP04,RB02}. Hence in
one direction, the proof is simply obtained by replacing the phase
gates with qubits measured with a non-Pauli angles, that are input
qubits. Then by Proposition~\ref{p-depth1}, the corresponding pattern
has depth~2.

To prove the other direction, let~$C$ be a circuit that can be
parallelized to a pattern $\mathcal P$ with depth~2. Hence from
Proposition~\ref{p-depth1} by adding  appropriate $(Z(\al))^{\dag}$
gates to the beginning of~$C$, we obtain another circuit~$C'$ that
translates to a pattern~$\mathcal P'$ with only Pauli measurements.
Now Theorem~4 in \cite{DKP04} implies that~$C'$ is in the
Clifford group and hence~$C$ has the desired form.
\end{proof}

A simple case of the above proposition is a Clifford circuit, that
was known already by \cite{Jozsa05,DKP04, RB02, RBB03}. On the other
hand, the best known result in terms of depth complexity for the
circuit implementing a subgroup of the Clifford group is
Proposition~\ref{p-moore} due to~\cite{MN02}. Using our
 forward and backward construction
of Section~\ref{s-cir-pattern},  we improve this and obtain a
general result on circuit depth complexity for the whole Clifford
group.

\PRO \label{p-clifford} Any quantum circuit on $n$ qubits of  size
$s \in \poly(n)$ consisting of Clifford gates can be parallelized to
a circuit with~$O(\log n)$ depth and~$O(s^3 +n )$  auxiliary qubits.
\ORP

Hence from Propositions~\ref{p-cdepth2} and~\ref{p-clifford}, we see
a logarithmic improvement in depth for implementations in the MBQC
compared to the circuit model. What we achieve actually is a
translation of quantum logarithmic depth in a circuit to constant
quantum depth plus classical logarithmic  depth in a pattern. We now
show that this separation is tight by giving an example of a unitary
that can be implemented as a pattern with constant quantum depth,
but that \emph{must} have logarithmic depth in the quantum circuit
model.

\begin{lemma}
Let $U_p$ be the parity unitary transformation defined by
\AR{U_p\,\ket{x_1, x_2, \ldots , x_n} = \ket{x_1, x_2,\ldots, x_{n-1}, \bigoplus_{i=1}^n x_i}\,.
}
Assume~$C$ to be any circuit consisting of 1-
and 2-qubit gates that implements this unitary. Then the depth
of~$C$ is in~$\Omega(\log n)$.
\end{lemma}

\begin{proof}
Since the state of the last output qubit depends on every input
qubit, and the circuit has only 1-- and 2--qubit gates,  the depth
of the circuit must be in~$\Omega(\log n)$.
\end{proof}

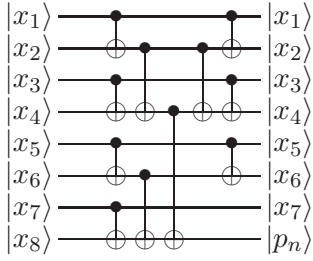
\begin{figure}
\[
\Qcircuit @C=1em @R=1.1em {
& \ket{x_1}& & \qw & \ctrl{1}   & \qw       & \qw         & \qw  & \ctrl{1}  & \qw & \ket{x_1}\\
& \ket{x_2}& &\qw & \mytarg     & \ctrl{2}  & \qw          & \ctrl{2} & \mytarg  & \qw & \ket{x_2}\\
& \ket{x_3}& &\qw & \ctrl{1}    & \qw       &\qw           & \qw& \ctrl{1} & \qw & \ket{x_3}\\
& \ket{x_4}& &\qw & \mytarg     &  \mytarg  & \ctrl{4}    & \mytarg & \mytarg& \qw & \ket{x_4}\\
& \ket{x_5}& &\qw & \ctrl{1}    & \qw       & \qw          & \qw  & \ctrl{1} & \qw & \ket{x_5}\\
& \ket{x_6}& &\qw & \mytarg     & \ctrl{2}  & \qw          & \qw   & \mytarg& \qw & \ket{x_6}\\
& \ket{x_7}& &\qw & \ctrl{1}    & \qw       & \qw           & \qw &\qw & \qw & \ket{x_7} \\
& \ket{x_8}& &\qw & \mytarg     & \mytarg   & \mytarg      &\qw &
\qw & \qw & \ket{p_n} }
 \] \caption{A logarithmic-depth circuit for parity unitary transformation, where $p=\bigoplus_{i=1} x_i$.} \label{fig:Parity circuit}
\end{figure}

Figure~\ref{fig:Parity circuit} gives a logarithmic depth circuit
for~$U_p$. This circuit uses only Clifford gates and hence
by Proposition~\ref{p-cdepth2}, we can implement it as a pattern
with depth~2. Note however that the pattern has a classical
logarithmic depth, which reconciles the depths in the two models:
the \emph{sum} of the classical and quantum depths in the pattern is
equal to the total quantum depth in the circuit.

\section{Discussions and future directions}

The design of parallel algorithms is one of the main challenges in both
classical and quantum computing and has  significant impact on theory
and implementations.
The advantage of quantum computing
models over classical counterparts has been extensively studied in the context of computational complexity, whereas
relatively little is known in terms of depth complexity.
In addition, the comparison of models of quantum computing has
been mainly explored from the computational  aspect although other
measures of comparison such as parallelism might lead to new
directions in our understanding of the power and limitations of
quantum computing.

In this paper, we considered  two well-known models of quantum
computing, the circuit model and the measurement-based quantum
computing, and presented a logarithmic separation between them in
terms of quantum depth complexity. We further demonstrated how a
simple forward and backward transformation between circuits and
measurement patterns leads to an automated procedure of
parallelization. More importantly, the set of tools that we developed
to study the depth complexity, such as the notion of the
\emph{influencing paths}, result in a simple construction for
parallel patterns and circuits, this being the insertion of some
particular type of Clifford operation among the non-Clifford ones.

A simple way of observing the advantages of the MBQC over the quantum
circuit can be seen via the tradeoff between space and depth
complexity as the transformation from a circuit to MBQC adds some
auxiliary qubits and hence decreases the depth. On the other hand,
one can also argue that the advantage is due to a clear separation
of the types of depths that are involved in a computation: the
preparation, quantum computation and classical depths. In other
words, in the circuit model, all operations are done
``\emph{quantumly}'' whereas in a pattern, some part of the
computation can be performed via \emph{classical processing}.
This intuition seems to be also responsible for some of the
previously known results on circuit parallelization such as  the work of Robert Griffiths and Chi-Sheng Niu on the parallel semi-classical quantum Fourier
transform~\cite{GN96}. Hence it would be interesting to see if our
tools can indeed reproduce the same results for these or other classes of
circuits where the output qubits are always measured.

Although it is encouraging that we obtain a generic method
for circuit parallelization by exploiting the classical control
structure in MBQC, it is not clear at this stage how our set
of tools might be put in use to design parallel algorithms for a
given classical problem and further work in this direction is necessary.

Another direction to investigate is the extension of the
characterization results to the patterns with generalized
flow~\cite{g-flow}, a recently developed notion for MBQC computing
that provides both a necessary and sufficient condition for
determinism that might lead to a more parallel structure than patterns
with flow.

\section*{Acknowledgements}
We would like to thank Stephen Jordan for finding an error in
our earlier upper bound for classical depth and Iordanis Kerenidis for
his insightful comments on circuit lower bounds techniques.  We
further thank Sean Barrett, Hugue Blier, Richard Cleve, Vincent Danos and
Ge\v{n}a Hahn for insightful discussions.
A.\,B.'s work is partially supported by scholarships
from \textsc{Nserc}, \textsc{Fqrnt} and the \textsc{Cfuw}. E.\,K.'s
work is partially supported by \textsc{Aro-Dto}. We are grateful to
the \textsc{Ciar} for financing E.\,K.'s stay at the Universit\'e de
Montr\'eal where this collaboration began.

\appendix{}

\section{Introduction to quantum computing}
\label{Appendix-A}

Let $\mcl H$ denote a 2-dimensional complex vector space,
equipped with the standard inner product. We pick an orthonormal
basis for this space, label the two basis vectors
$\ket{0}$ and $\ket{1}$, and for simplicity identify them
with the vectors $\left(\begin{array}{c}1\\ 0\end{array}\right)$
and $\left(\begin{array}{c}0\\ 1\end{array}\right)$, respectively.
A \emph{qubit} is a unit length vector in this space, and so can be
expressed as a linear combination of the basis states:
$$
\alpha_0\ket{0}+\alpha_1\ket{1}=\left(\begin{array}{c}\alpha_0\\
\alpha_1 \end{array}\right).
$$
Here $\alpha_0,\alpha_1$ are complex \emph{amplitudes},
and $|\alpha_0|^2+|\alpha_1|^2=1$.

An \emph{$m$-qubit state} is a unit vector in the $m$-fold tensor space
$\mcl H\otimes\cdots\otimes \mcl H$.  The $2^m$ basis states of this space are
the $m$-fold tensor products of the states $\ket{0}$ and $\ket{1}$.
We abbreviate $\ket{1}\otimes\ket{0}$ to $\ket{1}\ket{0}$  or $\ket{10}$. With these basis states, an $m$-qubit state $\ket{\phi}$
is a $2^m$-dimensional complex unit vector
$$
\ket{\phi}=\sum_{i\in\{0,1\}^m}\alpha_i\ket{i}.
$$
There exists quantum states that cannot be written as the tensor
product of other quantum states, \eg~ $\ket {00} + \ket {11}$. This
means that given a general element of $\mcl H\otimes \mcl H'$ one
cannot produce elements of $\mcl H$ and $\mcl H'$; such states are
called \emph{entangled} states.

We use $\bra{\phi}=\ket{\phi}^*$ to denote the conjugate transpose
of the vector $\ket{\phi}$, and $\inp{\phi}{\psi}=\bra{\phi}\cdot\ket{\psi}$
for the inner product between states $\ket{\phi}$ and $\ket{\psi}$.
These two states are \emph{orthogonal} if $\inp{\phi}{\psi}=0$.
The \emph{norm} of $\ket{\phi}$ is $\norm{\phi}=\sqrt{|\inp{\phi}{\phi}|}$.

A quantum state can evolve by a unitary operation or by a
measurement. A \emph{unitary} transformation is a linear mapping that
preserves the norm of the states. If we apply a unitary $U$ to a state
$\ket{\phi}$, it evolves to $U\ket{\phi}$.

The \emph{Pauli operators} are a well-known set of unitary
transformations for quantum computing:
\begin{equation*} X= \begin{pmatrix} 0 & 1 \\
1 & 0
\end{pmatrix}, \,\,\, Y = \begin{pmatrix} 0 & -i \\ i & 0 \end{pmatrix}
, \,\,\, Z= \begin{pmatrix} 1& 0 \\0 & -1 \end{pmatrix}\, ,
\end{equation*}
and the \emph{Pauli group} on $n$ qubits is generated by Pauli
operators. Other well-known unitary transformations are the identity
$I$, the \emph{Hadamard} gate $H$, the \emph{phase} gate
$Z(\alpha)$, of which $Z(\pi/4)$ and $Z(\pi/2)$ are a special cases, and the
Controlled-Z gate $\ctR Z$:
\begin{equation*}
I = \begin{pmatrix} 1 & 0 \\
0 & 1
\end{pmatrix}, \,\,\, H= \frac{1}{\sqrt{2}}\begin{pmatrix} 1 & 1 \\ 1 & -1
\end{pmatrix},
\end{equation*}
\begin{equation*} Z(\alpha) = \begin{pmatrix} 1 & 0 \\ 0 &
e^{i\alpha}
\end{pmatrix}, 
\,\,\, \ctR Z =
\begin{pmatrix} 1 & 0 & 0 & 0   \\   0 & 1 & 0 & 0 \\ 0 & 0 & 1 & 0 \\0 & 0 & 0 & -1
\end{pmatrix}.
\end{equation*}

The \emph{Clifford group} on $n$ qubits is generated by the
following matrices: $Z, H, Z({\pi/2})$ and $\ctR Z$. This set of
matrices is not universal for quantum computation, but by adding any
single qubit gate not in the Clifford group (such as $Z({\pi/4})$),
we do get a set that is approximately universal for quantum
computing. The importance of the Clifford group for quantum
computation is that a computation consisting of only Clifford
operations on the computational basis followed by final Pauli
measurements (see below) can be efficiently simulated by a classical
computer, this is the Gottesman-Knill theorem
\cite{Gottesman97,NC00}.

The most general measurement allowed by quantum mechanics is
specified by a family of positive semidefinite operators
$E_i=M_i^*M_i$, $1\leq i\leq k$, subject to the condition that
$\sum_i E_i=I$. A projective measurement is defined in the special
case where the operators are projections. Let $\ket{\phi}$ be an
$m$-qubit state and $\mcl B=\{\ket{b_1},\ldots,\ket{b_{2^m}}\}$ an
orthonormal basis of the $m$-qubit space. A projective measurement
of the state $\ket{\phi}$ in the $\mcl B$ basis means that we apply
the projection operators $P_i=\ket{b_i}\bra{b_i}$ to $\ket{\phi}$.
The resulting quantum state is $\ket{b_i}$ with probability
$p_i=|\inp{\phi}{b_i}|^2$. An important class of projective
measurements are Pauli measurements, \ie~ projections to eigenstates
of Pauli operators.

\section{Introduction to the measurement-based model}
\label{Appendix-B}

The measurement-based model \cite{RB01, RB02, RBB03} is a relatively
new approach to quantum computation that is oriented around
single-qubit measurements and entanglement for performing quantum
computations. In this model, computations are represented as
\emph{patterns}, which are sequences of \emph{commands} acting on
the qubits in the pattern. These commands are of four types:
one-qubit preparations, two-qubit entanglement operations,
single-qubit measurements and single-qubit Pauli corrections. In
addition to this, there is a classical control mechanism, called
\emph{feed-forward}, that allows measurement and correction commands to
depend on the results of previous measurements.

This model contrasts with the widely-used approach to quantum
computing which is the quantum circuit model. In this model, qubits
are represented by wires, unitary operations are represented by
gates and measurements usually occur only at the end of the circuit,
their sole purpose being to obtain a classical output out of the
quantum output.

More precisely, here are the types of commands that are involved in a computation in the MBQC:
\begin{enumerate}
 \item  $N_i$ is a  one-qubit preparation command which prepares the
  auxiliary qubit~$i$ in state $\ket{+}= \frac{1}{\sqrt{2}} (\ket{0} +
  \ket{1})$. The preparation commands can be implicit from the
  pattern: when not specified, all non-input qubits are prepared in the $\ket{+}$
  state.
  \item $\et ij$ is a two-qubit entanglement command which applies the controlled-$Z$ operation,
  $\ctR Z$, to qubits $i$ and $j$. Note that the $\ctR Z$ operation is symmetric and so $E_{ij} =
 E_{ji}$. Also, $E_{ij}$ commutes with~$E_{jk}$ and so the ordering
 of the entanglement commands in not important.
 \item $\M\al i$ is a one-qubit measurement on qubit~$i$ which depends on parameter $\al \in [0,2\pi)$ called the
 \emph{angle of measurement}. $\M\al i$ is the orthogonal projection onto states
 \begin{align*}
 \ket{+_\al}&=\ost(\ket0+ e^{i\al}\ket1)\\
\ket{-_\al}&=\ost(\ket0-\ei\al\ket1),
\end{align*}
followed by a trace-out operator, since measurements are
destructive. We denote the classical outcome of a measurement done
at qubit $i$ by $s_i\in\ztwo$. We take the specific convention that
$s_i=0$ if the measurement outcome is~$\ket{+_\al}$, and
that~$s_i=1$ if the measurement outcome is~$\ket{-_\al}$. Outcomes
can be summed together resulting  in expressions of the form
\begin{equation*}
s=\sum_{i\in I} s_i
\end{equation*}
 which are called \emph{signals}, and where the
summation is understood as being done modulo~$2$.  The \emph{domain}
of a signal is the set of qubits on which it depends (in this
example, the domain of~$s$ is~$I$).

\item $X_i$ and $Z_i$ are one-qubit Pauli corrections which
correspond to the application of the Pauli $X$ and $Z$ matrices,
respectively,  on qubit $i$.
\end{enumerate}

In order to obtain universality, we have to add a feed-forward mechanism
which allows measurements and corrections to be dependent on the
results of previous measurements~\cite{RB01,DKP04}. Let~$s$ and~$t$
be signals. Dependent corrections are written as $\cx is$ and $\cz is$ and
dependent measurements are written as $\mLR \al ist $.  The meaning of
dependencies for corrections is straightforward: $\cx i0=\cz i0=I$
(no correction is applied), while $\cx i1=\Cx i$ and $\cz i1=\Cz i$\,.
In the case of dependent measurements, the measurement angle depends
on $s$, $t$ and $\al$ as follows:
\begin{align} \label{msem}
\mLR{\al} ist = \m{(-1)^s\al+t\pi} i
\end{align}
 so
that, depending on the parity of~$s$ and $t$, one
may have to modify the angle of measurement~$\al$ to one of $-\al$,
$\al+\pi$ and $-\al+\pi$. These modifications correspond to
conjugations of measurements under $X$ and $Z$:
\begin{align} \label{e-Xact}
\cx is\M{{\al}}i \cx is&=\M{{(-1)^s\al}}i\\
\label{e-Zact} \cz it\M{{\al}}i \cz it&=\M{{\al+t\pi}}i
\end{align}
 and so we will refer to them as the $X$- and
$Z$-actions or alternatively as the $X$- and $Z$-dependencies. Since
measurements are destructive, the above equations  simplify
 to:
\begin{align}
\M{{\al}}i \cx is&=\M{{(-1)^s\al}}i\label{xmx}\\
\M{{\al}}i \cz it&=\M{{\al+t\pi}}i\label{zmz}. \end{align}
Note that these two actions are commuting, since $-\al+\pi=-\al-\pi$ up to $2\pi$,
and hence the order in which one applies them doesn't matter.

A \emph{pattern} is defined by the  choice of a finite set $V$ of
qubits, two not necessarily disjoint sets $I \subseteq V$ and $O
\subseteq V$ determining the pattern inputs and outputs, and a
finite sequence of commands acting on $V$. We require that  no
command depend on an outcome not yet measured, that no command act
on a qubit already measured, that a qubit be measured if and only if
it is not an output qubit and that a qubit be prepared if and only if
it is not an input qubit. This set of rules is known as the
\emph{definiteness} condition.

Just as circuits, patterns operate on a fixed number of input
qubits. Such models of computation are called \emph{non-uniform}. If
we want to solve problems that are defined for an arbitrary input
length, we need to construct one pattern for each length. This
pattern family is an \emph{infinite} object. By imposing some
\emph{uniformity conditions}, we require that the patterns for
different input lengths have something in common concerning
their structure. This, in turn, ensures that a pattern family
has a finite description. These uniformity conditions
are similar to  those that are usually imposed on uniform families
of \emph{circuits}~\cite{Vollmer99}.

The execution of a pattern consists  in performing each command in
sequence, from right to left.  If $n$ is the number of measurements
(\ie~the number of non-output qubits), then this may follow $2^n$
different computational branches. Each branch is associated with a
unique binary string  $s$ of length $n$, representing the classical
outcomes of the measurements along that branch, and a unique
\emph{branch map} $A_{s}$ representing the linear transformation
from the input Hilbert space to the output Hilbert space, along that
branch.

A pattern is said to be \emph{deterministic} if all the branch maps
are proportional, it is said to be \emph{strongly deterministic}
when branch maps are equal (up to a global phase), and it is said to
be \emph{uniformly deterministic} if it is deterministic for any
choice of measurement angles. A pattern is said to be in
\emph{standard form} if all the preparation $N_i$ and entanglement
operators $\et ij$ appear first in its command sequence, followed by
measurements and finally corrections. A pattern that is not in
standard form is called a \emph{wild pattern}. Any  wild pattern can
be put in its unique standard form~\cite{DKP04}; this form can
reveal implicit parallelism in the computation, and is well-suited
for
 certain implementations~(see Section~\ref{sec:standardization and depth}).

The procedure of rewriting a pattern to its  standard form is called
\emph{standardization}. This can be done by applying the rewrite
rules~\eqref{e-EX}--\eqref{e-MZ}. The rewrite rules also contain the
\emph{free commutation rules} (Equations~\eqref{freecomm1}--\eqref{freecomm3}) which tell us that, if we are dealing
with disjoint sets of target qubits, measurement, corrections and
entanglement commands commute pairwise \cite{DKP04}.
\begin{align}
\label{freecomm1}\et ij\CO{\vec k}&\Rar\CO{\vec k}\et ij \quad\hbox{where $A$ is not an
  entanglement}\\
\CO{\vec k}\cx is&\Rar\cx is\CO{\vec k} \quad\hbox{where $A$ is not
a correction}\\
\label{freecomm3}\CO{\vec k}\cz is&\Rar\cz is\CO{\vec k} \quad\hbox{where $A$ is not a
  correction}
\end{align}
where~$\vec k$ represent the qubits acted upon by command~$A$,
and are  distinct from~$i$ and~$j$.  Clearly these rules
could be reversed since they hold as equations but we are orienting them
this way in order to obtain termination for the standardization procedure.

Under rewriting, the computation space, inputs and outputs remain
the same, and so do the entanglement commands.  Measurements might
be modified, but we still measure exactly the same qubits.  The only
major modifications concern local corrections and dependencies. If
there were no dependencies at the start, none will be created in the
rewriting process.

We can extend the  rewrite rules to include a command called
\emph{signal shifting}
(equations~\eqref{e-signal1}--\eqref{e-signal4}). This allows us to
dispose of dependencies induced by the $Z$-action, and obtain
sometimes standard patterns with smaller depth complexity (see
Section~\ref{sec:signalshifting and depth}).

\begin{equation*}
\overline{\,\,\,\,\,\,\,\,\,\,\,\,\,\,\,\,\,\,\,\,\,\,\,\,\,\,\,\,\,\,\,\,\,\,\,\,}
\end{equation*}


\begin{thebibliography}{10}

\bibitem{Pelli97}
T.~Pellizzari.
\newblock Quantum networking with optical fibres.
\newblock {\em Physical Review Letters}, 79:5242 -- 5245, 1997.

\bibitem{CZKM97}
J.{\,}I. Cirac, P.~Zoller, H.{\,}J. Kimble, and H.~Mabuchi.
\newblock Quantum state transfer and entanglement distribution among distant
  nodes in a quantum network.
\newblock {\em Physical Review Letters}, 78:3221--3224, 1997.

\bibitem{B00}
S.{\,}C. Benjamin.
\newblock Schemes for parallel quantum computation without local control of
  qubits.
\newblock {\em Physical Review~A}, 61:020301, 2000.

\bibitem{KLM01}
E.~Knill, R.~Laflamme, and G.{\,}J. Milburn.
\newblock A scheme for efficient quantum computation with linear optics.
\newblock {\em Nature}, 409:46--52, 2001.

\bibitem{Nielsen04}
M.{\,}A. Nielsen.
\newblock Optical quantum computation using cluster states.
\newblock {\em Physical Review Letters}, 93:040503, 2004.

\bibitem{BR04}
D.{\,}E. Browne and T.~Rudolph.
\newblock Resource-efficient linear optical quantum computation.
\newblock {\em Physical Review Letters}, 95:010501, 2005.

\bibitem{BK04}
S.{\,}D. Barrett and P.~Kok.
\newblock Efficient high-fidelity quantum computation using matter qubits and
  linear optics.
\newblock {\em Physical Review~A}, 71:060310(R), 2005.

\bibitem{Jozsa05}
R.~Jozsa.
\newblock An introduction to measurement based quantum computation.
\newblock Available as \url{http://arxiv.org/abs/quant-ph/0508124v2}, 2005.

\bibitem{GN96}
R.{\,}B. Griffiths and C.~Niu.
\newblock Semiclassical {F}ourier transform for quantum computation.
\newblock {\em Physical Review Letters}, 76:3228--3231, 1996.

\bibitem{CW00}
R.~Cleve and J.~Watrous.
\newblock Fast parallel circuits for the quantum {F}ourier transform.
\newblock In {\em Proceedings of the 41st Annual IEEE Symposium on Foundations
  of Computer Science (FOCS 2000)}, pages 526--536, 2000.

\bibitem{MN02}
C.~Moore and M.~Nilsson.
\newblock Parallel quantum computation and quantum codes.
\newblock {\em SIAM Journal on Computing}, 31:799–815, 2002.

\bibitem{GHP00}
F.~Green, S.~Homer, and C.~Pollett.
\newblock On the complexity of quantum {ACC}.
\newblock In {\em Proceedings of the 15th Annual IEEE Conference on
  Computational Complexity}, page 250, 2000.

\bibitem{GHMP02}
F.~Green, S.~Homer, C.~Moore, and C.~Pollett.
\newblock Counting, fanout and the complexity of quantum {ACC}.
\newblock {\em Quantum Information \& Computation}, 2:35--65, 2002.

\bibitem{RB01}
R.~Raussendorf and H.{\,}J. Briegel.
\newblock A one-way quantum computer.
\newblock {\em Physical Review Letters}, 86:5188--5191, 2001.

\bibitem{Nielsen05}
M.{\,}A. Nielsen.
\newblock Cluster-state quantum computation.
\newblock {\em Rep. Math. Phys.}, 57:147--161, 2006.

\bibitem{BB06}
D.{\,}E. Browne and H.{\,}J. Briegel.
\newblock One-way quantum computation --- a tutorial introduction.
\newblock Available as \url{http://arxiv.org/abs/quant-ph/0603226v2}, 2006.

\bibitem{DKP04}
V.~Danos, E.~Kashefi, and P.~Panangaden.
\newblock The measurement calculus.
\newblock To appear in \emph{Journal of the ACM}. Available as
  \texttt{http://arxiv.org/abs/0704.1263v1}.

\bibitem{RB02}
R.~Raussendorf and H.~Briegel.
\newblock Computational model underlying the one-way quantum computer.
\newblock {\em Quantum Information \& Computation}, 2:443--486, 2002.

\bibitem{dk05c}
V.~Danos and E.~Kashefi.
\newblock Determinism in the one-way model.
\newblock {\em Physical Review~A}, 74:052310, 2006.

\bibitem{Fey}
R.~Feynman.
\newblock Simulating physics with computers.
\newblock {\em International Journal of Theoretical Physics}, 21:467--488,
  1982.

\bibitem{Deutsch}
D.~Deutsch.
\newblock Quantum theory, the {Church-Turing} principle and the universal
  quantum computer.
\newblock {\em Proceedings of the Royal Society of London A}, 400:97--117,
  1985.

\bibitem{D89}
D.~Deutsch.
\newblock Quantum computational networks.
\newblock {\em Proceeding of the Royal Society of London A}, 425:73--90, 1989.

\bibitem{NC00}
M.{\,}A. Nielsen and I.{\,}L. Chuang.
\newblock {\em Quantum Computation and Quantum Information}.
\newblock Cambridge University Press, Cambridge, 2000.

\bibitem{graphstates}
M.~Hein, J.~Eisert, and H.{\,}J. Briegel.
\newblock Multi-party entanglement in graph states.
\newblock {\em Physical Review~A}, 69:062311, 2004.

\bibitem{Beaudrap06}
N.~de~Beaudrap.
\newblock Characterizing {\protect \&} constructing flows in the one-way
  measurement model in terms of disjoint {$\protect I$}--\;\!{$\protect O$}
  paths.
\newblock Available as \url{http://arxiv.org/abs/quant-ph/0603072v4}, 2006.

\bibitem{BDK06}
N.~de~Beaudrap, V.~Danos, and E.~Kashefi.
\newblock Phase map decomposition for unitaries.
\newblock Available as \url{http://arxiv.org/abs/quant-ph/0603266v1}, 2006.

\bibitem{MS04}
M.~Mhalla and S.~Perdrix.
\newblock Complexity of graph state preparation.
\newblock Available as \url{http://arxiv.org/abs/quant-ph/0412071v1}, 2004.

\bibitem{Diestel}
R.~Diestel.
\newblock {\em Graph Theory}.
\newblock Springer-Verlag, 2005.

\bibitem{g-flow}
D.{\,}E. Browne, E.~Kashefi, M.~Mhalla, and S.~Perdrix.
\newblock Generalized flow and determinism in measurement-based quantum
  computation.
\newblock Available as \url{http://arxiv.org/abs/quant-ph/0702212v1}, 2007.

\bibitem{Parity84}
M.~Furst, J.{\,}B. Saxe, and M.~Sipser.
\newblock Parity, circuits, and the polynomial-time hierarchy.
\newblock {\em Theory of Computing Systems}, 17:13--27, 1984.

\bibitem{generator04}
V.~Danos, E.~Kashefi, and P.~Panangaden.
\newblock Parsimonious and robust realizations of unitary maps in the one-way
  model.
\newblock {\em Physical Review~A}, 72:064301, 2005.

\bibitem{VC04}
F.~Verstraete and J.{\,}I. Cirac.
\newblock Valence-bond states for quantum computation.
\newblock {\em Physical Review~A}, 70:060302(R), 2004.

\bibitem{RBB03}
R.~Raussendorf, D.{\,}E. Browne, and H.{\,}J. Briegel.
\newblock Measurement-based quantum computation on cluster states.
\newblock {\em Physical Review~A}, 68:022312, 2003.

\bibitem{Gottesman97}
D.~Gottesman.
\newblock {\em Stabilizer codes and quantum error correction}.
\newblock PhD thesis, California Institute of Technology, 1997.

\bibitem{Vollmer99}
H.~Vollmer.
\newblock {\em Introduction to Circuit Complexity}.
\newblock Springer-Verlag, 1999.

\end{thebibliography}
\end{document}